\documentclass[pre,twocolumn,showkeys,amsmath,amssymb]{revtex4-1}
\usepackage[T1]{fontenc}
\usepackage{color}
\usepackage[latin1]{inputenc}
\usepackage{graphicx}
\usepackage{bbold}
\usepackage{epsfig}
\usepackage{rotating}
\usepackage{dcolumn}
\usepackage{bm}

\begin{document}

\title{Dynamical properties of heterogeneous nucleation of parallel hard squares}

\author{Miguel Gonz\'alez-Pinto}
\email{miguel.gonzalezp@uam.es}
\affiliation{Departamento de F\'{\i}sica Te\'orica de la Materia Condensada,
Universidad Aut\'onoma de Madrid,
E-28049, Madrid, Spain}
\author{Yuri Mart\'{\i}nez-Rat\'on}   
\email{yuri@math.uc3m.es}
\affiliation{Grupo Interdisciplinar de Sistemas Complejos (GISC), Departamento de Matem\'aticas, Escuela
Polit\'ecnica Superior, Universidad Carlos III de Madrid, Avenida de la Universidad 30, E-28911, Legan\'es, Madrid, Spain}
\author{Enrique Velasco}
\email{enrique.velasco@uam.es}
\affiliation{
Departamento de F\'{\i}sica Te\'orica de la Materia Condensada,
Instituto de F\'{\i}sica de la Materia Condensada (IFIMAC) and Instituto de Ciencia de Materiales Nicol\'as Cabrera,
Universidad Aut\'onoma de Madrid,
E-28049, Madrid, Spain}

\date{\today}

\begin{abstract}
We use the Dynamic Density-Functional Formalism and the Fundamental Measure Theory as applied to a fluid of 
parallel hard squares to study the dynamics of heterogeneous growth of non-uniform phases with columnar and
crystalline symmetries. The hard squares are (i) confined between soft repulsive walls with square symmetry, 
or (ii) exposed to 
external potentials that mimic the presence of obstacles with circular, square, rectangular or triangular 
symmetries. For the first case the final equilibrium profile of a well commensurated cavity consists of a 
crystal phase with highly localized particles in concentric square layers at 
the nodes of a slightly deformed square lattice. 
We characterize the growth dynamics of the crystal phase by quantifying the interlayer and intralayer fluxes and the non-monotonicity of the former, the saturation time, and other dynamical quantities. 
The interlayer fluxes are much more monotonic in time,
and dominant for poorly commensurated cavities, while the opposite is true 
for well commensurated cells: although smaller, the time evolution of interlayer fluxes are much more complex, presenting strongly damped oscillations which dramatically increase the saturation time. We also study how the geometry 
of the obstacle affects the symmetry of the final equilibrium non-uniform phase
(columnar vs. crystal). For obstacles with fourfold symmetry,
(circular and square) the crystal is more stable, while the columnar phase is stabilized for 
obstacles without this symmetry (rectangular or triangular). We find that,
 in general, density waves of columnar 
symmetry grow from the obstacle. However, additional particle localization along the wavefronts gives rise to a crystalline structure 
which is conserved for circular and square obstacles, but destroyed for the other two obstacles where columnar symmetry is restored.
\end{abstract}


\maketitle

\section{Introduction}
\label{intro}

The Dynamic Density Functional Theory (DDFT) has proved, since its first derivation 
in Ref. \cite{umberto1}, to be a very useful tool to extend the study of soft matter 
systems from equilibrium to non-equilibrium situations. The response of colloidal systems to  
time dependent, in general inhomogeneous, external fields has been extensively studied within this 
formalism \cite{Tarazona1,Tarazona3,lowen7,lowen8}. The diffusion of vacancies through a crystalline structure 
\cite{Teeffelen}, 
the heterogeneous crystal nucleation \cite{lowen4,lowen6}, the dynamics of sedimentation processes 
\cite{Schmidt8,umberto3},
the diffusion  of colloidal spheres \cite{Roth1} or rods 
in nematics and smectics \cite{Dijkstra,Grelet}, 
and  the study of confined self-propelled rods \cite{lowen5}, are important examples of the variety of systems 
that were extensively studied within this theoretical tool. The orientational degrees of freedom of rods 
generate an additional complication in the numerical implementation of DDFT, which can be avoided
by resorting to the restricted-orientation (Zwanzig) approximation \cite{Oettel}. We should bear in mind that this formalism 
was derived from the stochastic Langevin dynamics of Brownian particles  
in the overdamped limit \cite{umberto1}, and some caution should be taken to use it in far-from-equilibrium situations.
In general, the relaxation to the equilibrium dynamics is reasonably well described by DDFT.     

By construction, better performance of DDFT is obtained when the system at equilibrium is well described 
by an approximate grand-canonical free-energy density functional (DF), the main ingredient of DDFT. Recent work has extended
the DDFT by using a canonical DF (extracted from the grand-canonical one), which is 
more appropriate for systems with fixed number of particles \cite{Heras}. 
As is well known, the DFs with the highest performance are those for hard particle interactions, such 
as hard rods in 1D \cite{Percus} (whose DF is known exactly), parallel hard 
squares (PHS) \cite{Cuesta1,Cuesta2} or hard disks (HD) \cite{disks} in 2D, and hard spheres (HS) \cite{Roth2} in 3D, 
all of them based on the original FMT proposed by Rosenfeld \cite{Rosenfeld1}. 
Some coarse-grained DFs, such as those based on phase-field-crystal models, are obtained from microscopic DFs by an appropriate 
order-parameter gradient expansion. These models were successfully used to study the dynamical properties 
 of heterogeneous crystallization in monolayers of paramagnetic colloidal spheres \cite{lowen2}. 
The phase-field-crystal approximation 
was also used to explore, through its numerically tractable implementation, all possible 
stable two- and three-dimensional liquid-crystal textures as a function of some parameters \cite{lowen2} 
describing particle interactions. There exist recent works on DDFT studies of fluids of HD and HS using accurate 
DFs \cite{Roth1,confined}. However these studies are scarce due to their complicated numerical implementation; 
in contrast, phase-field approximations are simpler due to the local dependence of the free-energy on the order parameters.  

The purpose of the present study is twofold. (i) We use an accurate DF, based on Fundamental-Measure Theory (FMT),
in combination with DDFT, to study the relaxation dynamics in fluids of PHS. The DF model used
\cite{Cuesta1,Cuesta2} has been tested at bulk and in highly confined situations \cite{miguel2}. Our
study extends the type of particle geometries (HD and HS) considered thus far. (ii) As shown below in this section, 
the FMT for PHS predicts the stability of columnar (C) and crystal (K) phases for particular density intervals. 
The present model may be used to understand how the dynamical properties of heterogeneous nucleation induced by external 
potentials depend on the degree of commensuration between the C or K lattice parameters and the characteristic lengths 
of the confining external potentials. We are interested in the full dynamics,
from the initial to the final equilibrium states. For some external potentials and thermodynamic 
conditions, the system can be dynamically arrested in metastable states for very long times. We  characterize the 
dynamics of confined PHS inside square cavities in terms of relaxation times and of
properties of the interlayer and intralayer fluxes such as non-monotonicity, maximum values, etc. 
The heterogeneous nucleation of C and K phases from obstacles with different geometries are also studied. 
The symmetry of the growing phase crucially depends on the geometry and size of obstacles, and in fact for
carefully selected obstacles an unstable phase at bulk can nucleate.
  
\begin{figure}
\epsfig{file=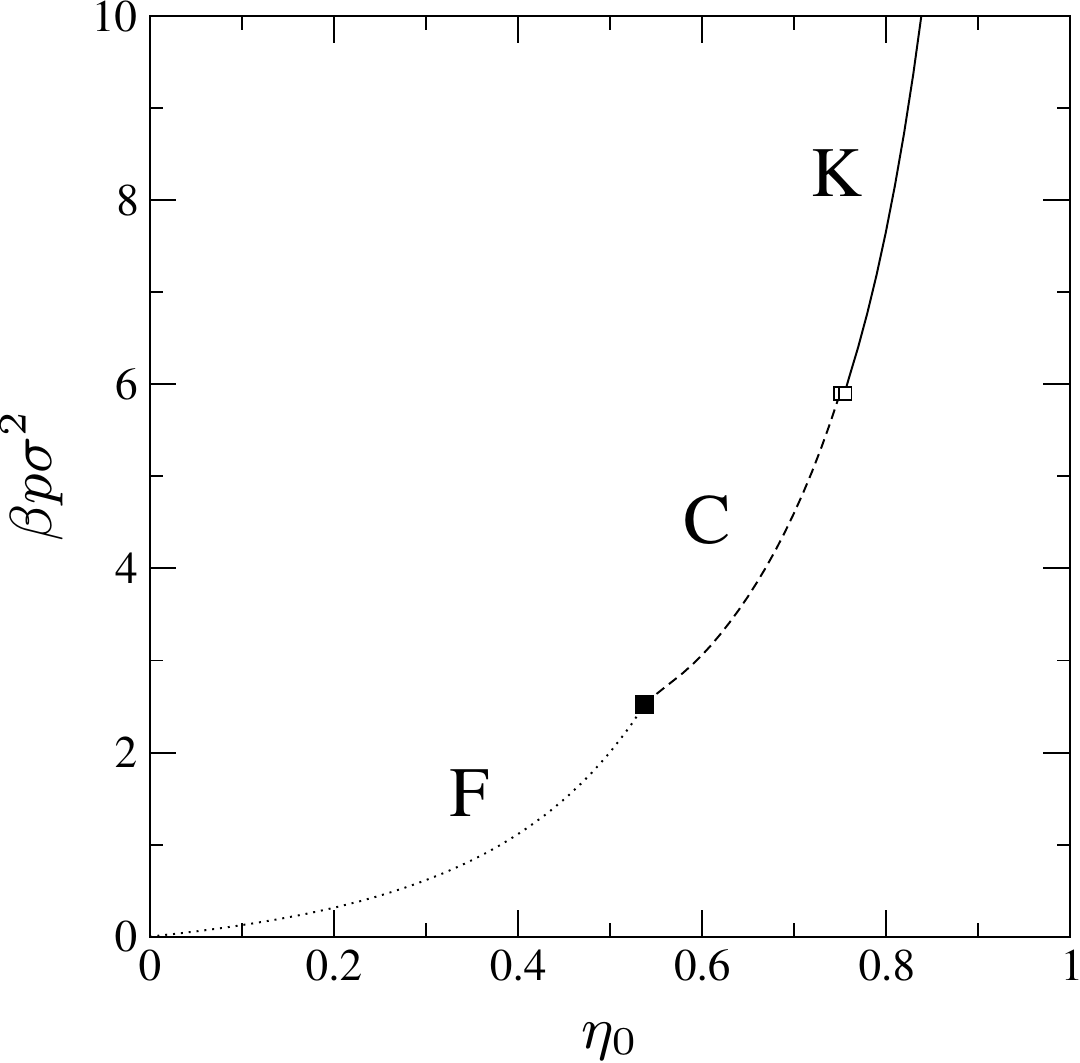,width=2.5in}
\caption{Equation of state (pressure in reduced units vs. mean packing fraction) of 
PHS from FMT.}
\label{eos}
\end{figure}

\begin{figure}
\epsfig{file=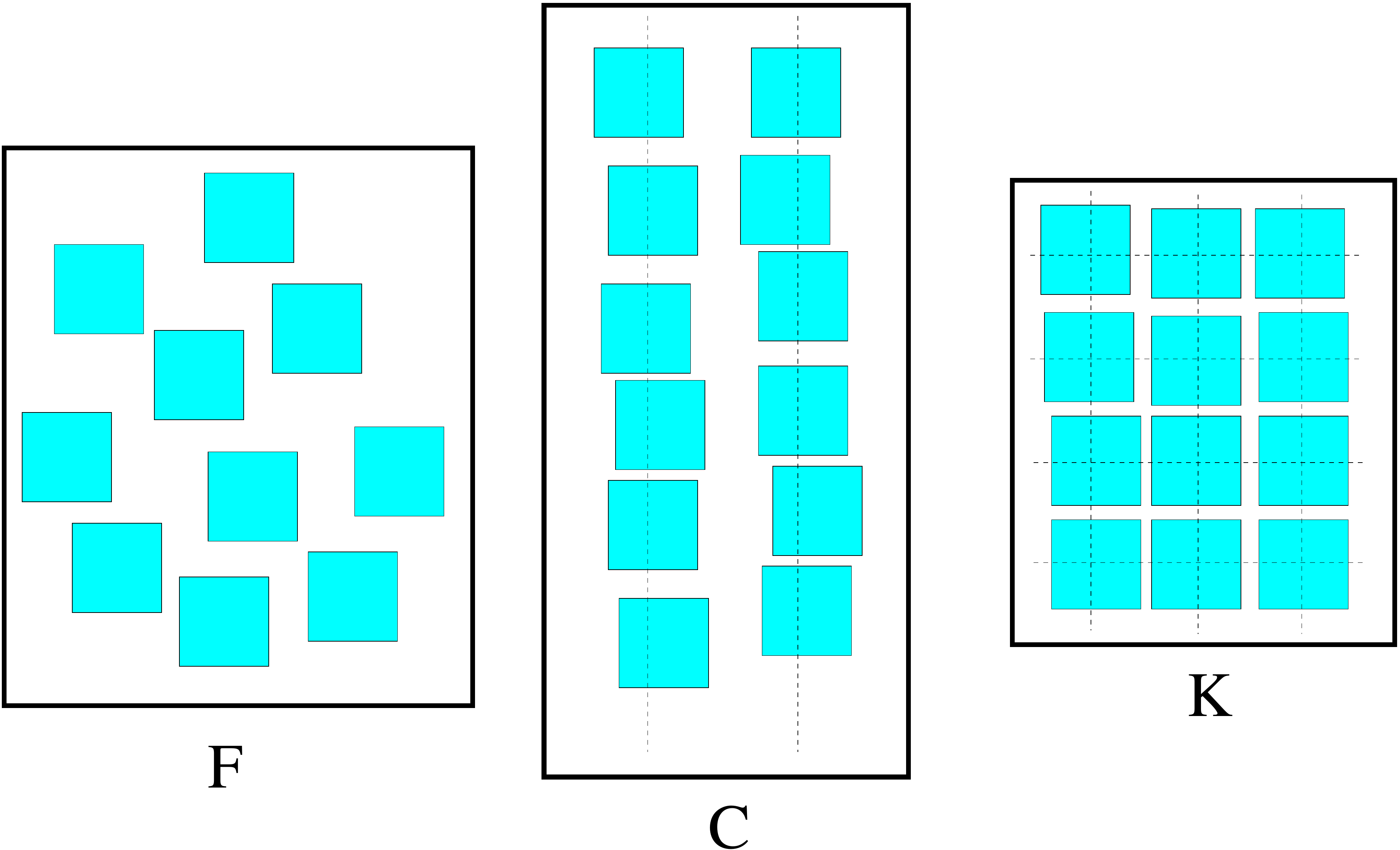,width=2.5in}
\caption{(Color online). Sketch of F, C and K phases (from left to right) that the FMT of PHS predicts.}  
\label{sketch}
\end{figure}

The FMT theory applied to a fluid of PHS predicts the equation of state (EOS) shown 
in Fig. \ref{eos}. The fluid (F) phase is stable up to a mean packing fraction $\eta_0=\rho_0\sigma^2$ 
($\sigma$ is the side length of the squares) equal to 0.534, at which a second-order transition to a columnar (C) phase 
takes place. The latter is stable up to $\eta_0\simeq 0.73$ (from free DF minimization \cite{roij})
or $\eta_0\simeq 0.75$ (from a Gaussian density-profile parameterization \cite{miguel2}). For higher densities
a crystalline phase (K) with simple square symmetry is stable up to close packing. 
See Fig. \ref{sketch} for a sketch of the different stable phases.

The article is organized as follows. In Sec. \ref{model} we present the model and the FMT-based DF which is used to
implement DDFT. We specify in Sec. \ref{ext_init} the external potential and the initial conditions used to study 
crystallization induced by confinement, while in Sec. \ref{magnitudes} we define the quantities that characterize the dynamics. 
In Sec. \ref{results} results for the crystallization of PHS induced by confinement are presented. This section is in turn divided 
into three parts devoted to different initial conditions used:
uniform (Sec. \ref{constant}), C (Sec. \ref{C_evolution}) and K (Sec. \ref{K_evolution}) density profiles as 
initial conditions.  Sec. \ref{obstacle} is concerned with the study of heterogeneous nucleation of C and K phases induced by the 
presence of obstacles of different sizes and symmetries. Finally some conclusions are drawn in Sec. \ref{conclusions}.

\section{Model}
\label{model}
The relaxation dynamics to equilibrium is studied using the
DDFT formalism of Ref. \cite{umberto1},
\begin{eqnarray}
\frac{\partial \rho}{\partial t}({\bm r},t)=-\boldsymbol{\nabla}\cdot {\bm J}({\bm r},t),
\label{umberto}
\end{eqnarray}
where $\rho({\bm r},t)$ is the local density.
The local flux, ${\bm J}({\bm r},t)$, is defined by
\begin{eqnarray}
{\bm J}({\bm r},t)=-{\cal D}\rho({\bm r},t)
\boldsymbol{\nabla}\frac{\delta \beta {\cal F}[\rho]}
{\delta \rho({\bm r},t)},
\label{umberto1}
\end{eqnarray}
where ${\cal D}$ is the diffusion constant, and
\begin{eqnarray}
\beta {\cal F}[\rho]=\int d{\bm r} 
\left[\Phi({\bm r},t)+\rho({\bm r},t)\beta V_{\rm ext}({\bm r},t)\right],
\end{eqnarray}
is the free-energy DF. $\beta=1/k_BT$ is 
the inverse temperature, $V_{\rm ext}({\bm r},t)$ is
the confining external potential, and $\Phi({\bm r},t)=\Phi_{\rm id}({\bm r},t)+
\Phi_{\rm ex}({\bm r},t)$ is the free-energy density, which is split in ideal
\begin{eqnarray}
\Phi_{\rm id}({\bm r},t)=\rho({\bm r},t)\left\{\log{\left[\rho({\bm r},t)
\Lambda^2\right]}-1\right\},
\end{eqnarray}
and excess 
\begin{eqnarray}
\Phi_{\rm ex}({\bm r},t)=-n_0({\bm r},t)\log\left[1-n_2({\bm r},t)\right]
+\frac{n_{1x}({\bm r},t)n_{1y}({\bm r},t)}{1-n_2({\bm r},t)},
\nonumber\\
\end{eqnarray}
parts. $\Lambda$ is the thermal length. The excess part 
corresponds to PHS \cite{Cuesta1,Cuesta2}, and the weighted densities
$n_{\alpha}({\bm r},t)=\int d{\bm r}' \rho({\bm r}',t)
\omega^{(\alpha)}({\bm r}-{\bm r}')$, 
are convolutions of the density profile and the following one-particle weighting functions: 
\begin{eqnarray}
&& \omega^{(0)}({\bm r})=\frac{1}{4}\delta\left(\frac{\sigma}{2}-|x|\right)
\delta\left(\frac{\sigma}{2}-|y|\right), \\
&& \omega^{(1x)}({\bm r})=\frac{1}{2}\Theta\left(\frac{\sigma}{2}-|x|\right)
\delta\left(\frac{\sigma}{2}-|y|\right),\\
&& \omega^{(1y)}({\bm r})=\frac{1}{2}\delta\left(\frac{\sigma}{2}-|x|\right)
\Theta\left(\frac{\sigma}{2}-|y|\right),\\
&& \omega^{(2)}({\bm r})=\Theta\left(\frac{\sigma}{2}-|x|\right)
\Theta\left(\frac{\sigma}{2}-|y|\right).
\end{eqnarray}
$\delta(x)$ and $\Theta(x)$ are the Dirac-delta and Heaviside functions, respectively.

\subsection{External potential and initial conditions}
\label{ext_init}

\begin{figure}
\epsfig{file=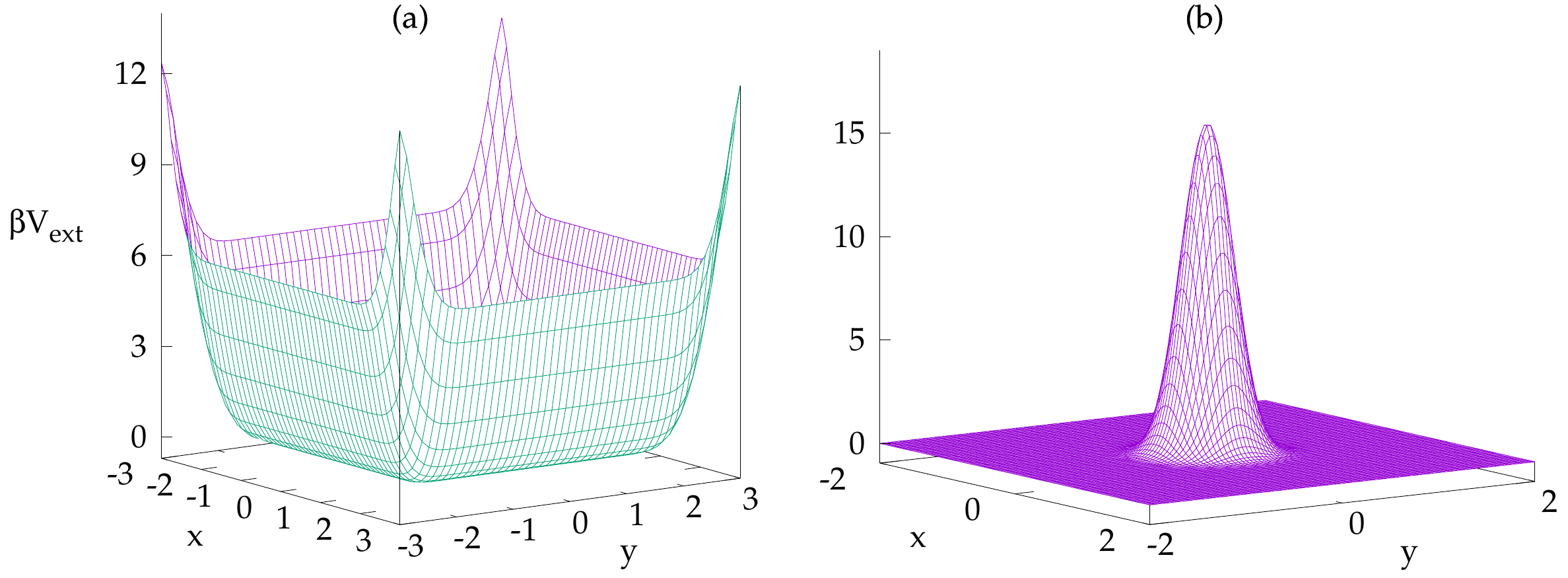,width=3.5in}
\caption{(Color online). Form of the external potentials defined by Eqs. (\ref{ext}) (a) 
and (\ref{potencial_2}) (b).}
\label{fig1_ext}
\end{figure}

Our first study concerns the dynamic evolution to equilibrium of a fluid of 
confined PHS when the confining external potential is switched on at $t=0$. 
The potential is defined in a region 
$x\in[-\frac{h}{2},\frac{h}{2}]$, $y\in[-\frac{h}{2},\frac{h}{2}]$,
where $h$ is the side of the square cavity. The following form is used:
\begin{eqnarray}
&&\beta V_{\rm ext}({\bm r},t)=-\epsilon \log\left\{\left[1-\Psi(x)\right]\times \left[
1-\Psi(y)\right]\right\}\Theta(t),\label{ext}
\end{eqnarray}
where
\begin{eqnarray}
&&\Psi(z-h/2)=\frac{1}{2}\sum_{k=-N}^{N}\left\{
\text{erf}\left[\sqrt{\alpha}\left(z+\frac{\sigma}{2}+kh\right)\right]\right.\nonumber\\
&&\left.-\text{erf}\left[\sqrt{\alpha}\left(z-\frac{\sigma}{2}+kh\right)\right]\right\},
\quad z=\{x,y\}, \label{sum}
\end{eqnarray}
with erf$(x)$
the standard error function. 
The external potential acts on the particles as a quickly decaying soft wall with characteristic 
inverse square length $\alpha$ and an amplitude $\epsilon$ that defines the height of the barrier.
For the sake of computational convenience, the box is periodically replicated, 
as defined by the sum over $k$ in Eqn. (\ref{sum}), forming a square lattice
of boxes. The number of boxes, dictated by the value of $N$,
is chosen large enough to guarantee the convergence of the sum (\ref{sum}) for 
$(x,y)\in [-\frac{h}{2},\frac{h}{2}]\times[-\frac{h}{2},\frac{h}{2}]$.
In Fig. \ref{fig1_ext}(a) 
the function $\beta V_{\rm ext}({\bm r},0)$ is plotted for 
some particular values of the parameters $\epsilon$, $\alpha$ and $h$.

In a second study we analyse the heterogeneous nucleation around obstacles with different geometries. 
For a circular obstacle, we define a repulsive external potential centred at ${\bm r}={\bm 0}$:
\begin{eqnarray}
\beta V_{\rm ext}({\bf r},t)=\epsilon \log\left[1+\Psi\left(\sqrt{x^2+y^2}\right)\right]\Theta(t), 
\label{potencial_2}
\end{eqnarray}
where
\begin{eqnarray}
\Psi(z)=\frac{1}{2}\left\{\text{erf}\left[\sqrt{\alpha}\left(z+D\right)\right]
-\text{erf}\left[\sqrt{\alpha}\left(z-D\right)\right]\right\}.\nonumber\\
\end{eqnarray}
Here $\epsilon$ and $\alpha$ have the same meaning as before, while $D$ is the characteristic 
dimension of the obstacle, in this case its diameter. In Fig. \ref{fig1_ext} (b) we plot the 
external potential for $D=1$ and certain values of the parameters $\{\epsilon,\alpha\}$. 
For a rectangular obstacle we use 
\begin{eqnarray}
\beta V_{\rm ext}({\bf r},t)=\epsilon \log\left[1+\Psi_x\left(x)\Psi_y(y)\right)\right]\Theta(t),
\end{eqnarray}
where
\begin{eqnarray}
\Psi_k(z)=\frac{1}{2}\left\{\text{erf}\left[\sqrt{\alpha}\left(z+D_k\right)\right]
-\text{erf}\left[\sqrt{\alpha}\left(z-D_k\right)\right]\right\}.\nonumber\\
\end{eqnarray}
$D_x$ and $D_y$ are the side-lengths of the rectangular obstacle along $x$ and $y$, respectively. The long,
$L=D_y$, and short, $D=D_x$, lengths of the rectangle will always be chosen to be parallel to the $y$ and $x$ axes, 
respectively. For $D_x=D_y=D$ we are describing a square obstacle.

It can be shown easily that the dynamic evolution that follows from Eqns. (\ref{umberto}) and 
(\ref{umberto1}) conserve 
the total number of particles, $N=\int_{A_{\rm cell}} d{\bm r} \rho({\bm r},t)$, 
where $A_{\rm cell}=[-\frac{h}{2},\frac{h}{2}]\times[-\frac{h}{2},\frac{h}{2}]$ 
is the area of the unit cell defined by the
external potential.  
Three different initial conditions were used for the density profiles: 
(i) $\rho({\bm r},0)=\rho_0$, i.e. a uniform density profile,  
(ii) $\rho({\bm r},0)=\rho_0^{(\rm C)}(y)$, corresponding to the bulk equilibrium 
density profile of columnar (C) symmetry,
and (iii) $\rho({\bm r},0)=\rho_0^{(\rm K)}({\bm r})$, corresponding to the scaled bulk 
equilibrium density profile of crystalline (K) symmetry. Both density profiles 
were previously calculated by fixing the mean number density $\rho_0$ (obtained from
integration of the density profile over the unit cell) to those values for which 
these phases are stable or metastable at bulk. By scaled density profile we mean 
an equilibrium density profile scaled along the $x$ and $y$ directions so as to
commensurate with the unit cell of the external potential, multiplied by a corresponding 
factor to obtain the same mean number density $\rho_0=A_{\rm cell}^{-1}\int_{A_{\rm cell}}d{\bm r}
\rho({\bm r},0)$. 

\begin{figure*}
\epsfig{file=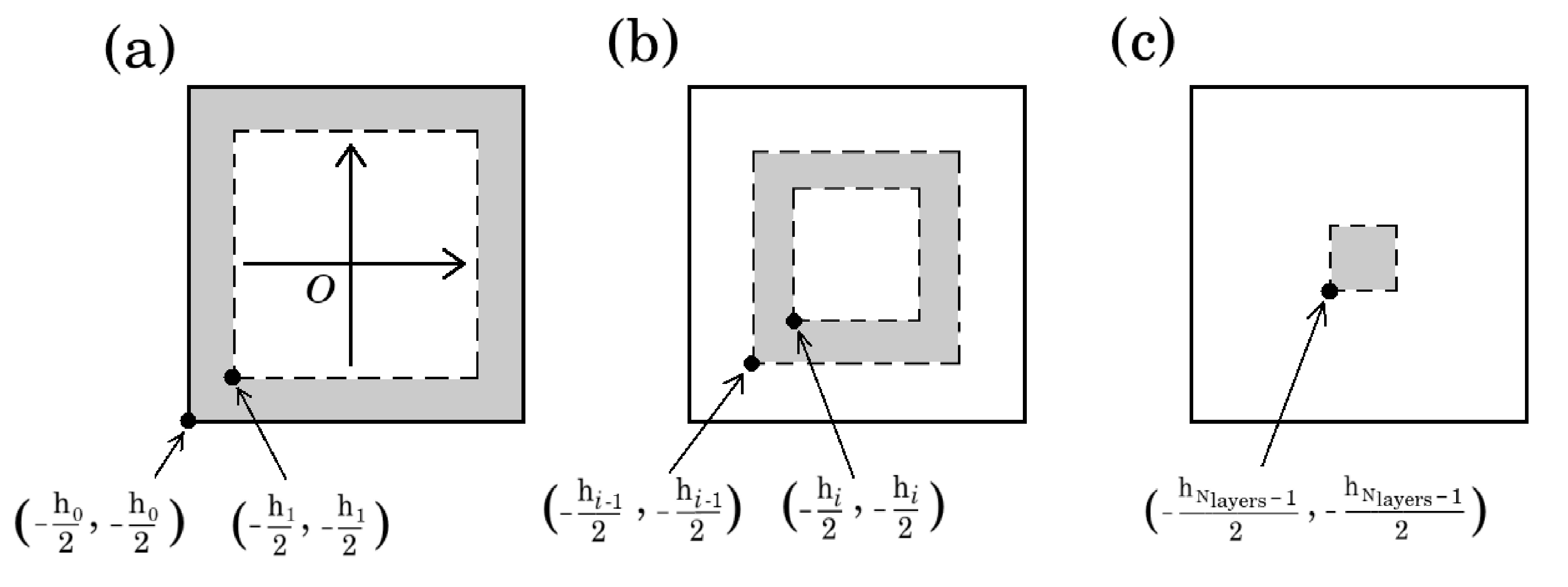,width=6.5in}
\caption{Schematic of the square rings defined. (a) First layer. (b) 
Intermediate layer. (c) Central layer.}
\label{esquema}
\end{figure*}

\subsection{Quantities to characterize the dynamics}
\label{magnitudes}
In this section we define the different quantities that characterize the relaxation dynamics. 
As shown in Sec. \ref{results} the final equilibrium state of the confined system 
consists of well-localized density peaks positioned in concentric 
square-like chains. We define a layer $A_i$ ($i=1,\dots,{\cal N}_{\rm layers}$) 
as a square ring containing each of the above 
chains and with boundaries defined by joining the local minima 
of $\rho_{\rm eq}({\bm r})$ between neighbor chains. The minima are 
located on the lines defined by
\begin{eqnarray}
&&\left(\pm\frac{h_j}{2},y\right),\hspace{0.4cm}-\frac{h_j}{2}\le y\le\frac{h_j}{2},\nonumber\\
&&\left(x,\pm\frac{h_j}{2}\right),\hspace{0.4cm}-\frac{h_j}{2}\le x\le\frac{h_j}{2},
\end{eqnarray}
where $j=0,\cdots,{\cal N}$ (with ${\cal N}=N_{\rm layers}$ for 
$N_{\rm layers}$ even, and ${\cal N}=N_{\rm layers}-1$ for $N_{\rm layers}$ odd),
while $h_0=h$ and $h_{{\cal N}_{\rm layers}}=0$.
The first layer is located 
next to the soft walls, while the last layer is at the centre.
The $i$th layer is then defined by the region (see Fig. \ref{esquema})
\begin{eqnarray}
\frac{h_{i}}{2}\le\left|x\right|\le\frac{h_{i-1}}{2},\hspace{0.6cm}
\frac{h_{i}}{2}\le\left|y\right|\le\frac{h_{i-1}}{2}.
\end{eqnarray} 
The innermost chain consists of either a single particle or 
four particles, depending on whether the total number of layers is an odd or an even number, respectively. 

The total particle flux across the boundaries of $A_i$ (the {\it interlayer} flux) 
is in turn equal to minus the exchange rate in number of particles, $-N_i'(t)$, inside $A_i$,
as can be shown by integrating Eq. (\ref{umberto}) over $A_i$:
\begin{eqnarray}
&&{\cal J}_i^{(\rm inter)}(t)\equiv -N_i'(t)=-\frac{d}{dt}\int_{A_i} d{\bm r}\rho({\bf r},t)
\nonumber\\&&=
\int_{A_i} d{\bm r} \boldsymbol{\nabla}
\cdot {\bm J}({\bm r},t)
=\int_{{\cal L}_i} dl {\bm J}({\bm r},t)\cdot {\bm n}_i.
\label{flux}
\end{eqnarray} 
The last term in (\ref{flux}), obtained from Gauss theorem, 
is a line integral over the boundary ${\cal L}_i$ of the $i$th layer 
with outer normal ${\bm n}_i$. 
Note that the total number of particles is a conserved quantity, so that
$\displaystyle\sum_i {\cal J}_i^{(\rm inter)}=0$.
Using the square symmetry, it is easy to show that the line integral can be computed as
\begin{eqnarray}
&&N_i'(t)=8\left[\int_0^{h_i/2}dy J_x\left(\frac{h_i}{2},y\right)\right.\nonumber\\
&&\left.-\int_0^{h_{i-1}/2}dy J_x\left(\frac{h_{i-1}}{2},y\right)\right],\nonumber\\
\end{eqnarray}
where we have defined 
\begin{eqnarray}
J_x(a,y)=-\left.\rho(x,y)\frac{\partial}{\partial x}\left(
\frac{ \delta \beta {\cal F}[\rho]}{\delta\rho(x,y)}\right)\right|_{x=a}.
\end{eqnarray}

We define the saturation time as a time $T_{\rm sat}$ such that
\begin{eqnarray}
T_{\rm sat}=t:\ 
\sum_i \left|{\cal J}^{(\rm inter)}_i(t)\right|<\delta, 
\end{eqnarray}
where $\delta$ is a tolerance (to be defined below).
The total interlayer flux over the whole cell $A_{\rm cell}=\cup_i A_i$ and integrated over time
is defined as 
\begin{eqnarray}
{\cal J}^{(\rm inter)}=\sqrt{\frac{1}{\tau_B}
\sum_i \int_0^{T_{\rm sat}} dt \left[N_i'(t)\right]^2}, 
\end{eqnarray}
where $\tau_B=\sigma^2/{\cal D}$ is the Brownian time,
while the maximum value of the interlayer fluxes over the whole cell and time is quantified through
\begin{eqnarray}
{\cal M}^{(\rm inter)}=\text{max}_{i,t} \left(|N_i'(t)|\right).
\end{eqnarray}
The non-monotonicity of the interlayer fluxes is taken into account by counting the total 
number of extrema of $\{N_i'(t)\}$ as a function of time:
\begin{eqnarray}
{\cal E}^{(\rm inter)}=\sum_i \#\text{extrema}\left[N_i'(t)\right],
\end{eqnarray}
Another useful quantity, measuring the total flux in the cell during the
complete time evolution, is
\begin{eqnarray}
{\cal J}^{(\rm total)}=\frac{1}{\tau_B\sigma}\int_0^{T_{\rm sat}} dt
\int_{A_{\rm cell}}d{\bm r} \left[\left| J_x({\bm r},t)\right|
+\left| J_y({\bm r},t)\right|\right],\nonumber\\
\end{eqnarray}
with $J_{x,y}({\bm r},t)$ the $x$ and $y$ components of ${\bm J}({\bm r},t)$.

To characterize the equilibrium density profiles we use apart from the total number of layers, 
${\cal N}_{\rm layers}$, the value of the highest density peak over the whole cell:
\begin{eqnarray}
\rho_{\rm max}=\text{max}_{{\bm r}\in A_{\rm cell}}\left[\rho_{\rm eq}({\bm r})\right].
\end{eqnarray}
Finally we define the mean packing fraction of the layer $i$ as 
\begin{eqnarray}
\eta_i(t)=A_i^{-1}\int_{A_i} d{\bm r} \rho({\bm r},t)\sigma^2,
\end{eqnarray}
Note that, as the areas $A_i$ ($i=1,\dots,{\cal N}_{\rm layers}$) are in general different, 
we have that
\begin{eqnarray}
&&\frac{1}{\cal N}_{\rm layers}
\sum_{i=1}^{{\cal N}_{\rm layers}} \eta_i(t)\nonumber\\
&&\neq \eta_0\equiv A_{\rm cell}^{-1}
\int_{A_{\rm cell}} d{\bm r} \rho({\bm r},t) \sigma^2=\rho_0\sigma^2,
\end{eqnarray}
i.e. the average of the mean packing fractions per layer is not a conserved 
quantity and it is 
different from the total mean packing fraction $\eta_0$, which is conserved.  

\section{Crystallization induced by confinement}
\label{results}

This section is devoted to the study of the dynamical relaxation of the 
confined fluid to equilibrium from different initial conditions. 
In Sec. \ref{constant} we present the 
results obtained from constant-density initial conditions,
while in Secs. \ref{C_evolution} and \ref{K_evolution} 
initial conditions with C and K symmetries are respectively chosen.

First we discuss an important issue on the terminology used in the article 
to describe the dynamic evolution of the density profile. We use sentences
like "particles are expelled from the walls" or 
"particles are highly localized/delocalized". With this we mean that the 
structure of the density profile is strongly changing with time: 
density peaks get smeared out or sharpened in space. One should always
bear in mind that there is no direct relation between a single density peak 
and a real particle, since density profiles measure the probability density 
of finding a particle at some particular position. The spatial integral 
of the density profile over a region with the same particle dimensions gives 
the probability to find the particle at this position and obviously this
can be less than one even for the K phase due to the existence of vacancies. 
We decided to keep this terminology for simplicity, avoiding the use of an 
excessively elaborate language.   

\subsection{Dynamic evolution from a constant density profile}
\label{constant}

\begin{figure}
\epsfig{file=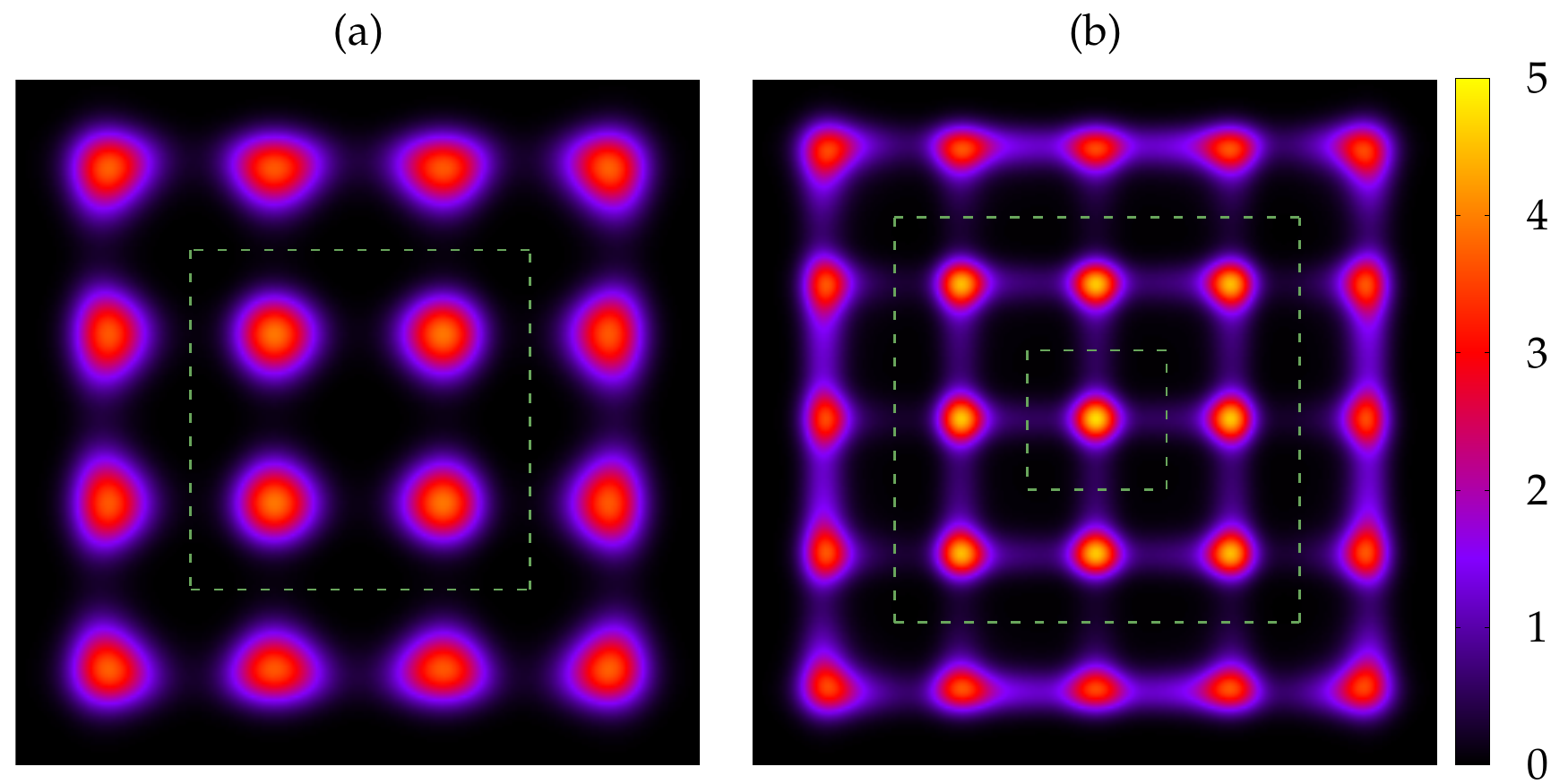,width=3.5in}
\caption{(Color online). Equilibrium density profiles 
$\rho_{\rm eq}^*({\bf r})$ inside the square cells of dimensions 
$h/\sigma=5.1$ (a) and 5.8 (b) starting from constant density profiles corresponding to packing fraction 
$\eta_0=0.6$. They are shown through false color contour plot images with 
colour scales correspondingly shown. Green dashed lines represent the boundaries 
between different layers [with a total amount of two (a) and three (b) layers].}
\label{fig0}
\end{figure}

We use a simple iteration scheme to solve Eqn. (\ref{umberto}):
the density profile at the $n$th timestep 
$t_n=n\Delta t$ is calculated from the previous one as 
\begin{eqnarray}
&&\rho^*(x_i,y_j,n+1)=\rho^*(x_i,y_j,n)\nonumber\\+&&\Delta\tau \boldsymbol{\nabla}\cdot 
\left(\rho^*(x_i,y_j,n)\boldsymbol{\nabla}
\left.\frac{\delta\beta{\cal F}[\rho]}{\delta\rho}\right|_{x_i,y_j,n}\right),
\nonumber\\
&& \{i,j\}=[0,M], \ n\geq 0
\label{convergence}
\end{eqnarray}
where $\Delta\tau=\Delta t/\tau_B$
and $\Delta t$ is the timestep. Also $\rho^*=\rho\sigma^2$, and the 
variables $\{x_i,y_j\}$ are the $x$ and $y$ 
coordinates of a node on the square 
grid used to discretize the cell, $[-\frac{h}{2},\frac{h}{2}]\times[-\frac{h}{2},\frac{h}{2}]$ 
($M\Delta x=h$, with $\Delta x=\Delta y=
\sigma/40$ the size of the spatial grid). 
The spatial derivatives in (\ref{convergence}) 
were calculated using a central finite-difference method. 
As a first study, we have chosen the initial local packing fraction
$\rho(x_i,y_j,0)\sigma^2=\eta_0=0.6$ $\forall \ \{i,j\}$, i.e. 
the initial density profile inside the cell is 
constant (and different from the bulk value, as can be seen 
from Fig. \ref{eos} which shows a stable C phase at $\eta_0=0.6$). 
Fig. \ref{fig0} presents the equilibrium density profiles after convergence 
of Eqn. (\ref{convergence}) at $T_{\rm sat}=n_{\rm sat}\Delta t$ for cells of 
dimensions (a) $h/\sigma=5.1$ and (b) $h/\sigma=5.8$, respectively.
The value of $\delta$ used to define convergence was $\delta\cdot\tau_B=10^{-5}$.

Despite the fact that the C phase is stable at bulk, 
the confining external potential localizes particles at the nodes
of a simple square lattice. The lattice parameter $a$ and the cell 
dimension $h$ are approximately related by
$h\simeq 2{\cal N}_{\rm layers}a$ or $h\simeq 
\left(2{\cal N}_{\rm layers}-1\right)a$ when the number of 
layers, $N_{\rm layers}$, is an even or an odd integer, respectively. 
For the latter case case a central peak 
is always found at the centre of the cell. For $h/\sigma=5.8$ the density peaks are sharper and more localized than those corresponding 
to $h/\sigma=5.1$, which are smeared out over space. 
This is a consequence of the difference between the lattice parameters 
of the confined system, $a$, and that 
of the metastable K phase at bulk, $a_{\rm K}$. 
The C phase is stable for packing fractions in the interval
$\eta_0\in[0.534,0.73]$. However, a metastable free-energy branch of K phase 
also bifurcates from the F branch at $\eta_0=0.534$, its free energy being above
the C branch until they cross at
$\eta_0\simeq 0.73$. When $a/a_{\rm K}\sim 1$ highly localized peaks are
present in the cavity, as shown in Fig. \ref{fig0} (b); 
otherwise the density profile is similar to that of panel (a). 
When commensuration 
between $a$ and $a_{\rm K}$ is nearly perfect [as in (b)], the density 
profile develops bridges between neighbor particles   
belonging to the same layer (with boundaries indicated by green lines). 
This means that particle fluctuations along these directions are so favoured 
that the K phase can support a large fraction of vacancies.  

\begin{figure}
\epsfig{file=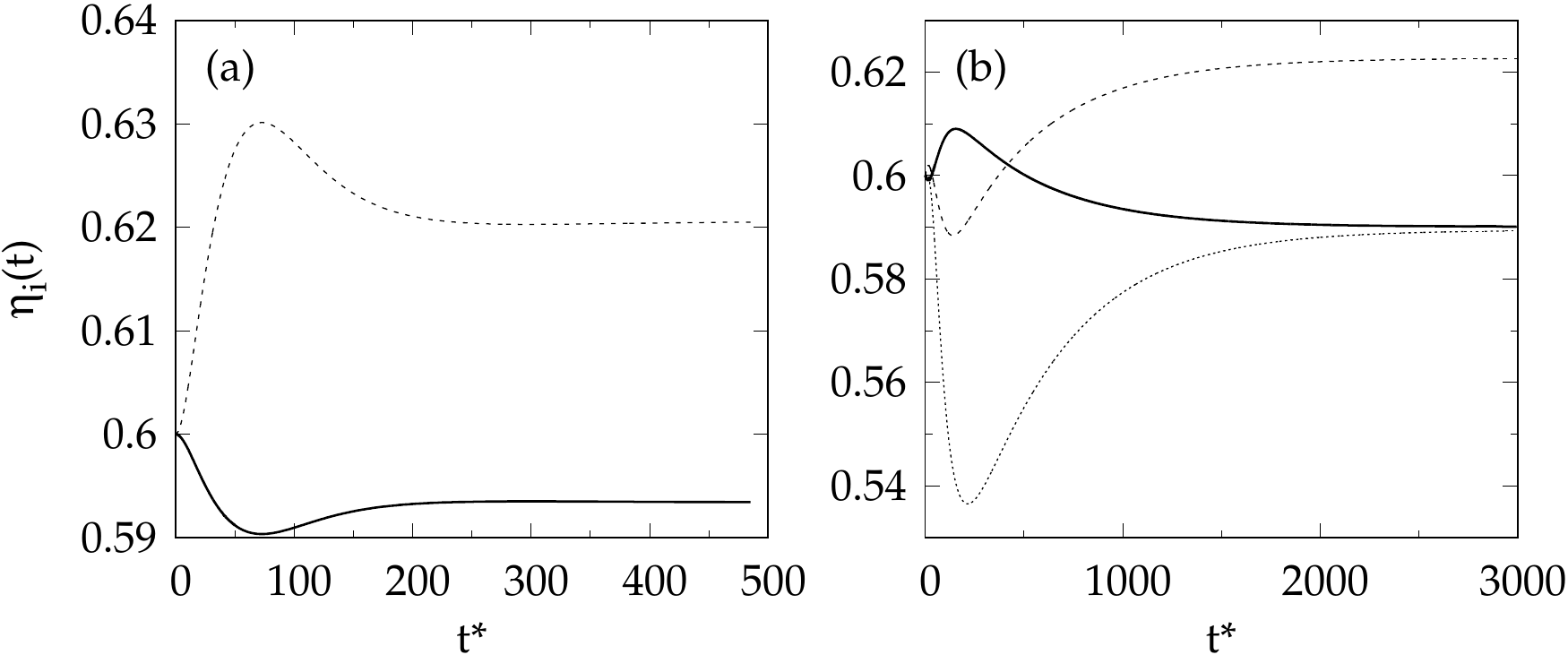,width=3.5in}
\caption{Mean packing fraction of layer $i$, $\eta_i(t)$, as a function of time, for square cells of dimensions 
$h/\sigma=5.1$ (a) and 5.8 (b). With solid, dashed and dotted lines are shown $\eta_1(t)$, $\eta_2(t)$ and 
$\eta_3(t)$ respectively. The initial condition corresponds to a constant density profile with $\eta_0=0.6$.}
\label{fig1}
\end{figure}

Fig. \ref{fig1} shows the dynamic evolution of the mean packing fraction 
of layer $i$, $\eta_i(t)$, as a function of scaled time $t^*=t/\tau_B=n\Delta\tau$
for the two cells shown 
in Fig. \ref{fig0}, which contain two ($h/\sigma=5.1$) 
and three ($h/\sigma=5.8$) layers, respectively. For the former, 
the first stages of the dynamic evolution of $\eta_1(t)$ present a small 
decrease, then a minimum and an increase to its stationary value 
$\eta_1(\infty)<0.6$, which is reached at $T_{\rm sat}^*\equiv T_{\rm sat}/\tau_B\simeq 500$. 
In this case the repulsive potential 
expels the excess of particles in contact with the soft wall, creating a 
first layer with lower mean packing fraction. 
By contrast the inner layer, formed at the end by four particles,
 increases its packing fraction, reaches a maximum, and tends to its stationary value 
$\eta_2(\infty)>0.6$. We can see that the dynamic evolution of the 
case $h/\sigma=5.8$ has the opposite behavior: the packing fraction of 
the first layer increases rapidly, reaches a maximum,
and finally decreases to a value $\eta_1(\infty)<0.6$, while the second 
layer exhibits the opposite evolution. Finally the third layer, enclosing 
at the end a single particle, exhibits the deepest minimum and a final 
relaxation to $\eta_3(\infty)<0.6$. As we will promptly  
see, cells that commensurate with the bulk lattice parameter, which
exhibit highly localized equilibrium density peaks [(b)], have intralayer 
fluxes which dominate over the interlayer ones, while
the opposite occurs when peaks are spatially smeared out, as in (a). 
Therefore the dominant effect of the external potential on the layers in 
(b) involves the motion of particles inside each layer to their equilibrium
highly localized positions and, in addition, the flow of particles to or from
the neighbour layers to make a regular square lattice.  
As a consequence of this complex dynamics, the saturation time is usually 
longer [$T_{\rm sat}^*\simeq 3000$ in (b), as compared with $500$ in (a)]. 
In contrast, for poorly commensurate cells [as in (a)], which give 
delocalized peaks, interlayer fluxes are more important and the dominant 
effect of the external potential on the first layer 
is to expel the excess of particles to the interior of the cell. The other 
layers get restructured by particle interchange with neighbour layers. 
The usual behavior in $\eta_2(t)$ is always opposite to that of 
$\eta_1(t)$ [see (a) and (b)]. Finally the third layer in (b), which
contains a single particle, reaches an equilibrium packing fraction less 
than $\eta_0$. Although these trends are generally true, there are exceptions 
to these behaviours, which can be explained by the inhomogeneities of the 
lattice parameter $a$ from the wall to the interior of the cell. 

\begin{figure}
\epsfig{file=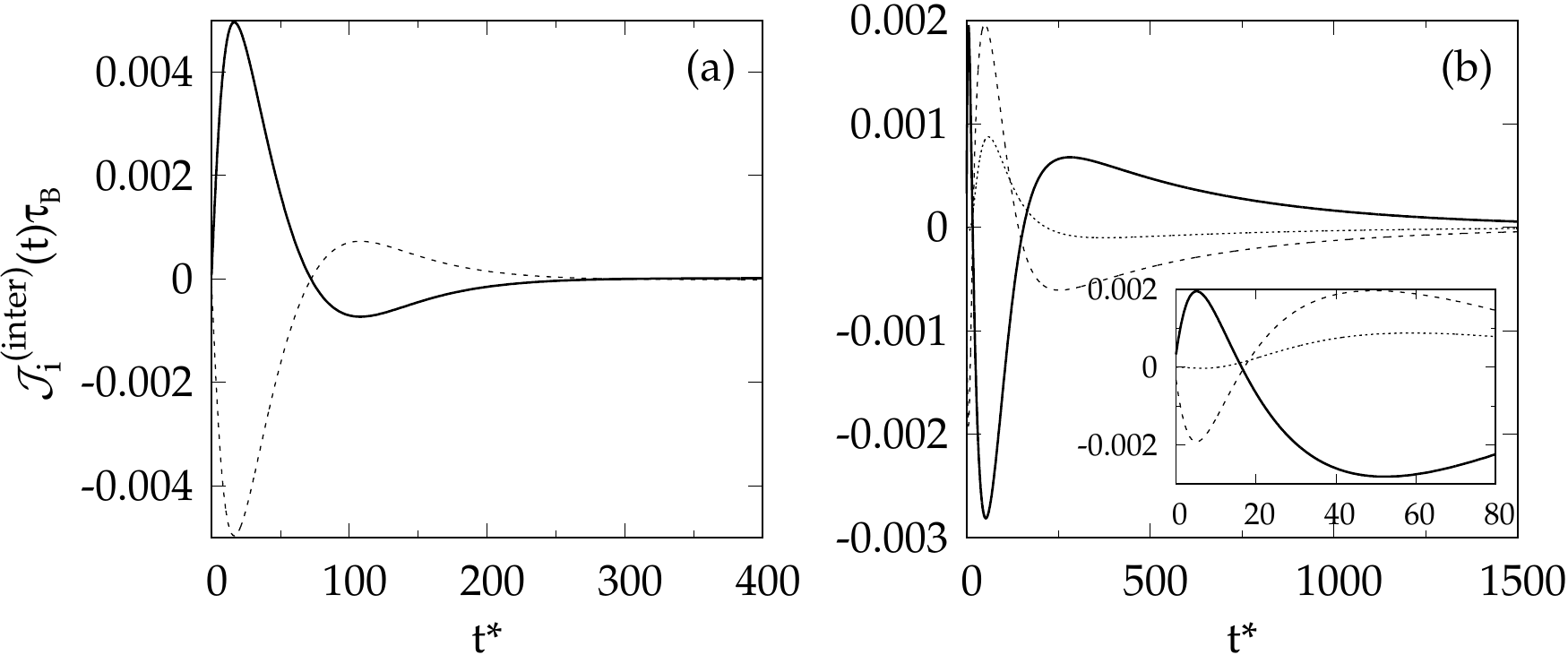,width=3.5in}
\caption{Interlayer fluxes, ${\cal J}^{(\rm inter)}_i(t)$, as a function of time, for square cells of dimensions 
$h/\sigma=5.1$ (a) and 5.8 (b). With solid, dashed and dotted lines are shown ${\cal J}^{\rm (inter)}_1(t)$, 
${\cal J}^{\rm (inter)}_2(t)$ and 
${\cal J}^{\rm (inter)}_3(t)$ respectively. The initial condition corresponds to a constant density profile 
with $\eta_0=0.6$.}
\label{fig2}
\end{figure}

The behaviour of the interlayer fluxes ${\cal J}_i^{(\rm inter)}(t)$ as a function of time 
confirms the preceding discussion. These are shown in Fig. \ref{fig2} for the same cells 
and initial conditions. For $h/\sigma=5.1$ the first cell becomes a source of particles,
creating a positive flux across its boundaries. This flux reaches a maximum, then decays
and reaches a minimum, and finally relaxes monotonically to the 
stationary state. Obviously the flux that crosses the boundaries of the second 
layer, ${\cal J}_2^{(\rm inter)}(t)$, is, by conservation of particles ($\displaystyle
\sum_i {\cal J}_i^{(\rm inter)}=0$), the specular reflection of 
${\cal J}_1^{(\rm inter)}(t)$ in the entire $t$-axis. The behaviour of the fluxes for 
$h/\sigma=5.8$ and $t^*>t_c\simeq 20$ is opposite to the previous case: particles
enter the first layer from the neighbor layer, so that ${\cal J}_1^{(\rm inter)}(t)$ 
becomes a negative, decreasing function down to a minimum, and then increases, changes 
sign at a certain time (the layer becoming a source of particles), reaches a maximum 
and finally relaxes to zero at a time $T_{\rm sat}$ much longer than in the previous case.   
The third, innermost layer, has a positive flux which relaxes to zero after 
reaching a small minimum, therefore becoming a source of particles. The above behavior 
pertains to times $t^*>t_c$. At very short times ($t^*<t_c$) the behavior is the opposite 
for the first two layers, and the same for the third (see inset). This latter fact 
confirms a scenario where the effect of the external potential propagates
from the walls to the inner layers with a finite velocity.
Although the extrema of the fluxes for $t^*<t_c$ 
and $t^*>t_c$ are of the same order
(see the inset), in the former case they are reached in very short times and, as a consequence,
the mean packing fractions $\eta_i$ have almost unnoticeable changes 
[see Fig. \ref{fig1}(b)]. Another important feature of fluxes in highly 
commensurate cavities, compared to noncommensurate ones, is the presence of a larger number 
of extrema [cf. (a) and (b)]. This is due to the fact that, as the front propagates
from the wall to the inner layers, intralayer fluxes --due to particle migration 
to their highly localized positions-- combined with outgoing and incoming fluxes from 
the neighbouring layers, result in nonmonotonic fluxes as the final equilibrium 
configuration is reached.

\begin{figure*}
\epsfig{file=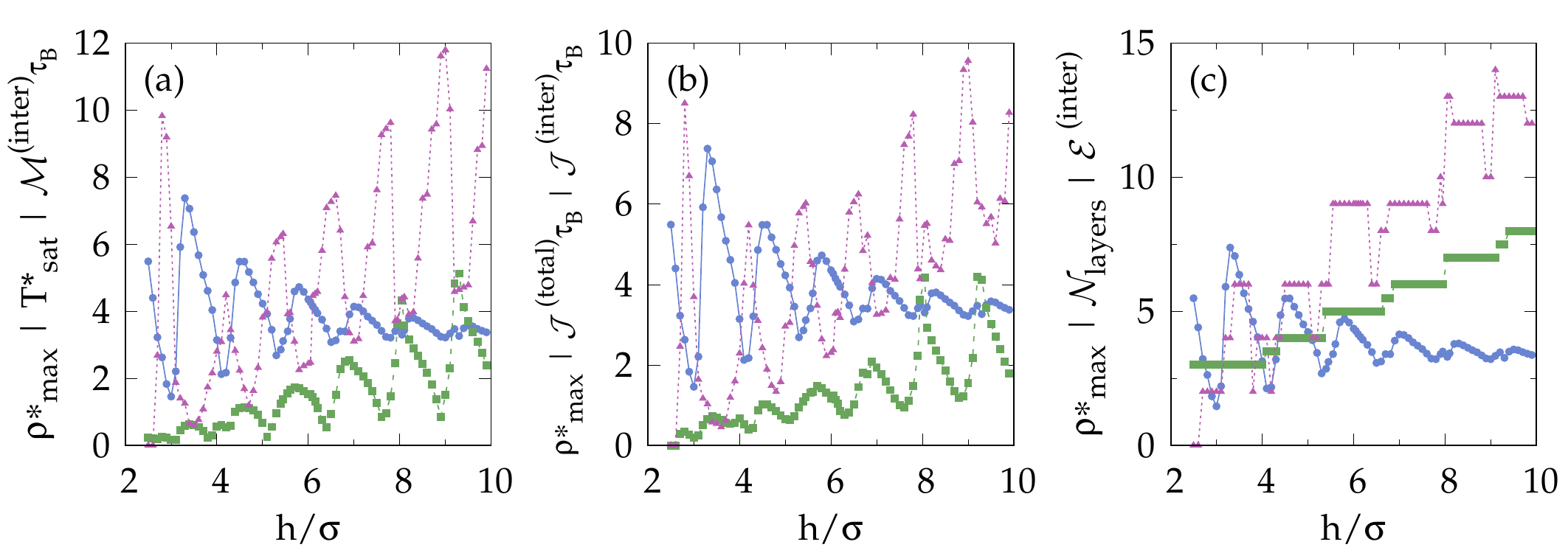,width=6.in}
\caption{(Color online). The maximum value of density peaks $\rho_{\rm max}^*$ [solid squares 
in (a), (b) and (c)], the scaled saturation time $T_{\rm sat}^*\times 10^{-3}$ [open circles in (a)], 
the scaled maximum value of the interlayer fluxes, 
${\cal M}^{(\rm inter)}\tau_B\times 8\cdot 10^2$ 
[gray triangles in (a)],
the scaled total interlayer flux ${\cal J}^{(\rm inter)}\tau_B\times 80$ [gray triangles in (b)], the scaled 
total flux 
${\cal J}^{(\rm total)}\tau_B\times 0.025$ [open circles in (b)], the number of extrema of interlayer 
fluxes ${\cal E}^{(\rm inter)}$ [gray triangles in (c)], 
and the total number of layers ${\cal N}_{\rm layers}$ 
[open circles in (c)] as a function of the cell dimension $h/\sigma$ corresponding to the dynamic 
evolution from a constant density initial profile with $\eta_0=0.6$.} 
\label{fig3}
\end{figure*}

Now we describe in detail the correlations between the different 
quantities (defined in Sec. \ref{magnitudes}) that characterize the dynamics 
as the cell dimension $h/\sigma$ is varied. In 
Fig. \ref{fig3}(a) we show the maximum of the equilibrium density profile at the cell, 
$\rho_{\rm max}$, the saturation time $T_{\rm sat}$, and the maximum value of the interlayer 
flux ${\cal M}^{(\rm inter)}$, as a function of $h/\sigma$. The saturation times $T_{\rm sat}$ 
are longer for well commensurate cells containing highly localized density peaks 
(maxima of $\rho_{\rm max}$). By contrast, the interlayer fluxes are less important: note
how the minima of ${\cal M}^{(\rm inter)}$ as a function of $h/\sigma$ are perfectly 
correlated with the maxima of $\rho_{\rm max}$. Therefore, (i) longer times are necessary 
to reach equilibrium states with highly structured density profiles, and (ii) particle
localization is dominated by intralayer, as opposed to interlayer, fluxes.

This scenario is clear from Fig. \ref{fig3}(b), where we can see that
the maxima of the total interlayer flux ${\cal J}^{(\rm inter)}$ correspond to poorly 
commensurate cells; equilibrium profiles with smeared out peaks are obtained by strong
interlayer fluxes where particles are exchanged between neighbouring layers. 
In contrast, well commensurate cavities reach their equilibrium states with much lower
values of ${\cal J}^{(\rm inter)}$. As total fluxes ${\cal J}^{(\rm total)}$  
are higher for well commensurate cavities [see panel (b)], while interlayer ones are less 
important, we can draw the important conclusion that intralayer fluxes are dominant
during relaxation to well structured density profiles. The nonmonotonicity 
of interlayer fluxes are well described by their total number of extrema 
${\cal E}^{(\rm inter)}$, and these is higher for well commensurate cells 
[see Fig. \ref{fig3} (c)]. The dynamic evolution from a constant density to a K phase 
with highly localized density peaks is more complex: particle migration
to well localized positions inside each layer with further restructuring through 
interlayer fluxes results in a highly nonmonotonic relaxation dynamics. 

Finally it is interesting to note that rapid changes in the total number of layers 
${\cal N}_{\rm layers}$ inside the cavity as $h/\sigma$ is changed take place for poorly 
commensurate cells with delocalized fluid-like density profiles [see Fig. \ref{fig3} (c)]. 
We have confirmed that the rapid change in ${\cal N}_{\rm layers}$ with $h/\sigma$, 
although related with the commensuration first-order transitions of a confined K phase 
inside a cavity with hard boundaries \cite{miguel3}, does not imply a phase transition. 
When hard walls are substituted by soft walls these transitions are suppressed.

\subsection{Dynamic evolution from a C density profile}
\label{C_evolution}

\begin{figure*}
\epsfig{file=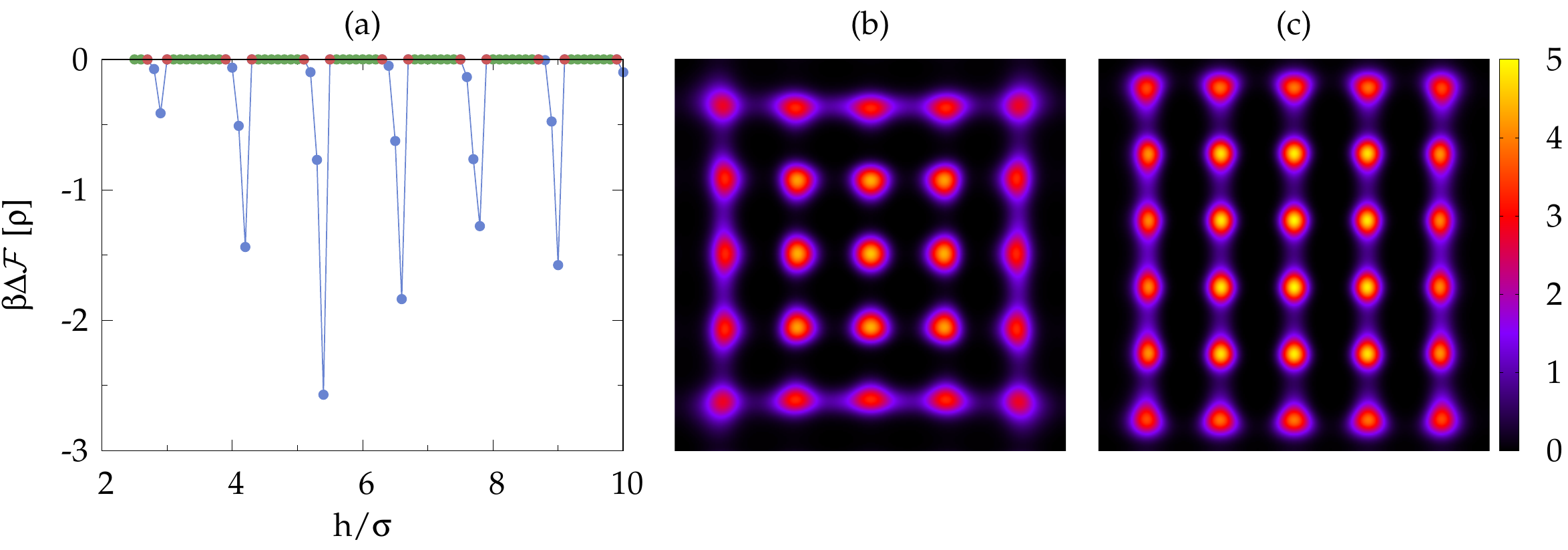,width=6.in}
\caption{(Color online). Free-energy differences [(a)] between asymmetric [(c)] and symmetric [(b)] finally converged 
density profiles. The latter are those corresponding to $h/\sigma=6.6$ 
resulting from the DDF evolution of 
confined PHS with initial density profiles corresponding to constant (b) and 
to bulk equilibrium C (c) density profiles, both having a mean packing fraction $\eta_0=0.57$.}
\label{arrested_1}
\end{figure*}

In this section we report on the differences between final states 
when the initial conditions are changed. For a cell with size $h/\sigma=6.6$,
we choose initial profiles corresponding to uniform density, Fig. \ref{arrested_1} (b), and
bulk equilibrium C phase, Fig. \ref{arrested_1} (c).
In the first case, panel (b), the final state is identical as before --a
symmetric K phase with layers formed by the same number of particles along $x$ and $y$ directions.
In the second, an asymmetric density profile is obtained, as shown in panel (c). 
Note that the number of particles in layers along the $y$ axis is one more 
than that along the $x$-axis. These asymmetric density profiles are 
always obtained when $h$ is very well commensurate with the lattice parameter, $a_{\rm C}$, of the C 
phase at bulk, i.e. when $h/a_{\rm C}\simeq k\in\mathbb{N}$. At this packing fraction the C phase 
is stable at bulk. If the cell size is such that an integer number 
of layers can be accommodated, then the total free energy will be lower than that 
of the symmetric density profile, panel (b). However the main effect of the external 
potential, as pointed out before, consists of the localization of particles at
the nodes of a regular lattice (of rectangular symmetry for asymmetric density profiles). 
Therefore, starting from a C density profile the system evolves by
keeping the same number of C layers along the $x$ 
direction (and consequently by fixing the lattice parameter along this direction to be $a_{\rm C}$), 
with a further localization of particles by diffusion along $y$ (parallel to the C layers) 
to their final positions. These positions are such that the lattice parameter is close to $a_{\rm K}$
(that of the metastable K phase at bulk) along the $y$ axis.

\begin{figure*}
\epsfig{file=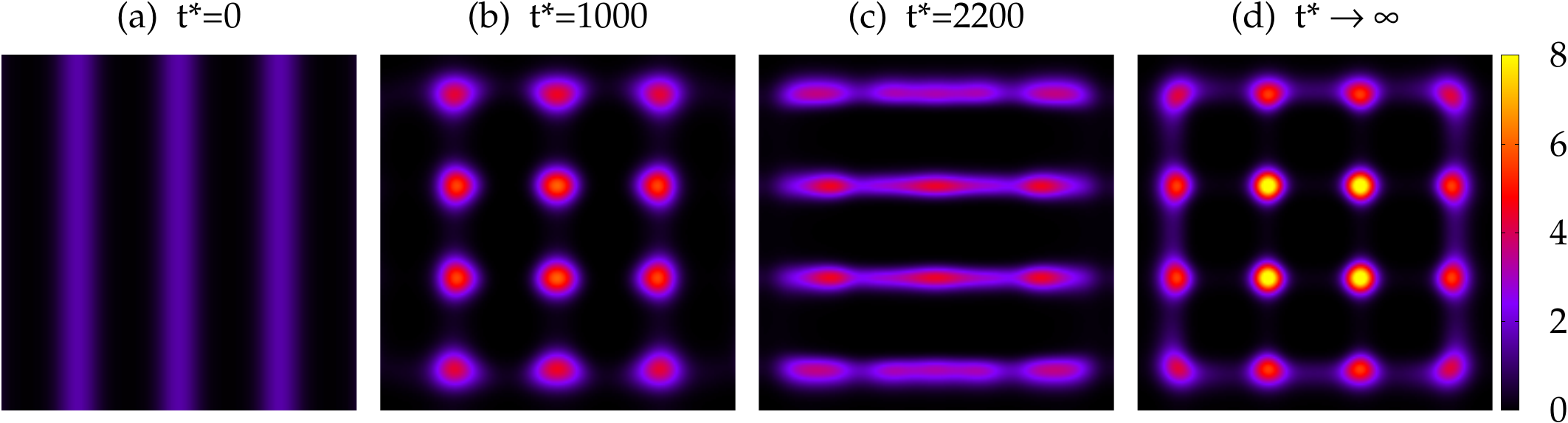,width=6.in}
\caption{(Color online). Density profiles corresponding to the dynamical evolution from a C initial profile 
at $t^*=0$ (a), $t^*=1000$ (b), $t^*=2200$ (c) and $t^*\to\infty$ (d) inside a cavity of 
$h/\sigma=4.3$.}
\label{arrested_2}
\end{figure*}

\begin{figure}
\epsfig{file=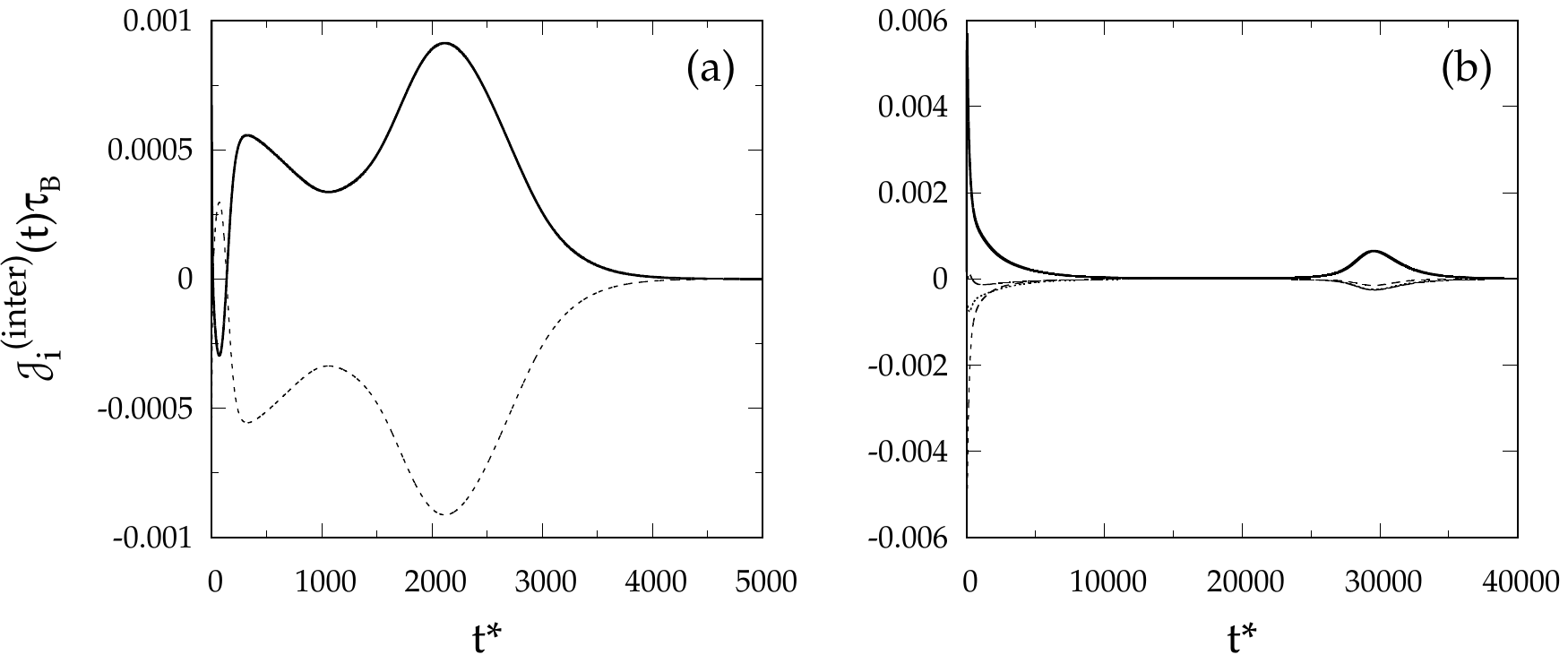,width=3.5in}
\caption{Interlayer fluxes corresponding to a dynamical evolution from C density profiles 
inside cavities of $h/\sigma=4.3$ (a) and 9.1 (b).}  
\label{time_evolution}
\end{figure}

Asymmetric density profiles, such as that in panel (c), are obtained only for special cells
that commensurate with $a_{\rm C}$. However, when this occurs, their free energies are
lower than that corresponding to the (metastable) K-symmetric profile [panel (b)]. 
This is shown in panel (a),  where the free-energy difference $\beta \Delta {\cal F}[\rho]
\equiv \beta\left({\cal F}\left[\rho^{(\rm asym)}\right]
-{\cal F}\left[\rho^{(\rm sym)}\right]\right)$ is plotted as a function of $h/\sigma$. The blue circles, corresponding to nonzero values, pertain to asymmetric density profiles, while the green triangles correspond to converged K-symmetric 
density profiles. Red squares indicate values of cavity size $h/\sigma$
for which a long-time dynamical evolution occurs; they are values 
with similar commensuration of $h/a_{\rm C}$ and $h/a_{\rm K}$, 
so that the system, depending on the initial conditions, 
could be arrested for a long time in metastable states. 
To illustrate this behaviour, Fig. \ref{arrested_2} shows density profiles 
at four different times. Panel (a), the initial condition, 
consists of a C density profile with three 
layers inside a cavity of $h/\sigma=4.3$. Fig. \ref{time_evolution}(a) shows 
the interlayer fluxes for the same system.    
As we can see from Fig. \ref{arrested_2}(b), the system initially evolves 
by localizing four different K peaks along each of the C layers, 
changing the density profile to an asymmetric K phase and selecting the 
distance between peaks along the $y$ direction to optimise the commensuration
with $a_{\rm K}$. This evolution occurs up to $t^*\sim 1000$ [see Fig. 
\ref{time_evolution} (a)]. However the free energy of this asymmetric metastable 
state is slightly above that corresponding to the $4\times 4$ symmetric 
K phase, and the system continues its evolution by further delocalizing 
the K peaks along $x$, creating four C layers parallel to this direction [see Fig. \ref{arrested_2} (c)]. This process lasts 
up to $t^*\sim 2000$ [see Fig. \ref{time_evolution} (a)] from which takes place the last 
dynamical path: the localization of four K peaks within each C layer to end 
in a $4\times 4$ symmetric K profile 
[see Fig. \ref{arrested_2} (d)]. Thus, we can conclude that, 
for some special values of $h/\sigma$,
the system can dynamically be trapped in metastable states
($3\times 4$ K profile for $h/\sigma=4.3$) during a long period of time 
($\sim 1000\tau_B$). For larger cavity sizes this effect is more dramatic, 
as can be seen in Fig. \ref{time_evolution}(b), 
where we show the interlayer fluxes corresponding 
to the dynamical evolution from a C phase with 7 layers up to the final equilibrium 
$8\times 8$ K profile inside a cavity of $h/\sigma=9.1$. 
We can see that the system is arrested into a $7\times 8$ K profile 
during $\sim 30000\tau_B$ after which the density profile is symmetrized through 
its columnarization along $x$ with a further localization of 8 K peaks along the columns to 
end in the symmetric $8\times 8$ density profile.
 

\subsection{Dynamic evolution from a K density profile}
\label{K_evolution}

\begin{figure*}
\epsfig{file=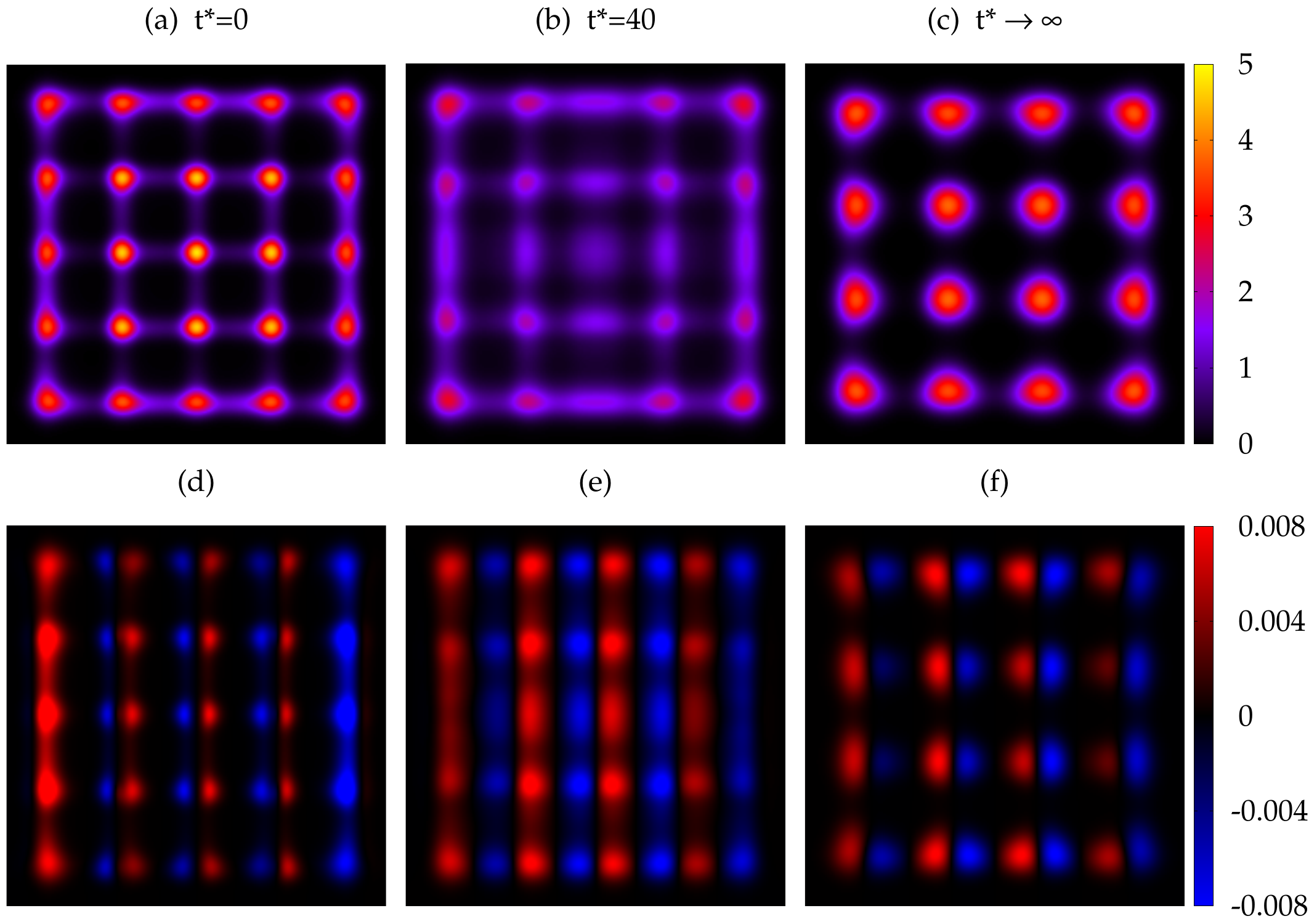,width=5.in}
\caption{(Color online). The initial (a), intermediate (b)  and finally converged (c) density profiles both 
with $\eta_0=0.6$. The initial density profile was conveniently scaled to fit it inside the 
cavity of $h/\sigma=5.1$. The fluxes $J_x({\bf r})$ are also plotted for times close to the initial 
(d), intermediate (e) and final (f) states. In (e) and (f) fluxes have been
multiplied by factors of 3 and 40, respectively.}
\label{evolving}
\end{figure*}

In preceding sections we described the dynamic evolution of confined PHS from F-like or C-like nonequilibrium 
initial conditions to their final states consisting of symmetric or asymmetric K-like density profiles. Now we 
proceed to describe the dynamics that follows our system departing from a non-equilibrium 
confined K-like symmetric density profile compressed enough that its total number of layers is one more than that 
corresponding to the equilibrium situation. The initial density profile was taken from the already 
converged density profile corresponding to  a wider cell and conveniently scaled along to $x$ and $y$ 
directions to fit it to the boundaries of the new cell. Also it is multiplied by a constant factor 
to fix to 0.6 the mean packing fraction over the cell. In Fig. \ref{evolving} we present the results 
corresponding to the cell of dimensions $h/\sigma=5.1$ and taking an initial density profile corresponding 
to the equilibrium one of a cavity with $h/\sigma=5.8$ properly scaled.   
The panels (a), (b) and (c) correspond to the initial ($t^*=0$), 
intermediate ($t^*=40$), and finally converged ($t^*\to\infty$) density 
profiles, while in (d), (e) and (f) panels we present the $x$-component of 
the local flux, $J_x({\bm r},t)$, for the same times.

We have found the following evolution from a three-layer density profile: 
(i) The density profile in the central square chains is delocalized over space, 
creating a smeared-out  
density profile along these directions, (ii) the rest of the peaks, even those corresponding to the 
most external (in contact with the soft wall), also delocalize along $x$ and $y$ 
directions and they move to the center of the cell creating an effective flux and 
(iii) the density profile is then restructured from the fluid-like density profile to the final one with only 
two, instead of three, layers and without any peak at the centre of the cell.  
This scenario is confirmed by the evolution of the fluxes: note in (e) how the highest values of the 
fluxes are located in the neighborhood 
of the central chains.  As we have already discussed above the identification of a peak as a particle 
could be misleading. 
The density profiles shown in (a) and (c) have a total amount 
of 25 and 16 peaks. However the mean 
packing fraction is the same ($\eta_0=0.6$) for both. This difference can be explained due to a
higher fraction of vacancies in the $4\times 4$ density profile. Note that if we approximately 
parameterize it as
\begin{eqnarray}
\rho({\bm r})\simeq
(1-\nu)\left(\frac{\alpha}{\pi}\right)^{3/2}
\sum_{{\bm R}_{\bm k}\in {\cal L}} e^{-\alpha\left({\bm r}-{\bm R}_{\bm k}\right)^2},
\end{eqnarray}
where $\nu$ is the fraction of vacancies, $\alpha$ is the Gaussian parameter which takes into account 
the extent of particle fluctuations around the positions ${\bm R}_{\bm k}$ of the square lattice ${\cal L}$, 
then the mean packing fraction can be approximately calculated as 
\begin{eqnarray}
\eta_0\simeq\frac{N_{\rm peaks}}{A_{\rm cell}}\int_{a_0}d{\bm r} \rho({\bm r})\sigma^2\simeq
\frac{N_{\rm peaks}(1-\nu)\sigma^2}{A_{\rm cell}}.  
\end{eqnarray}
with $a_0$ the unit cell containing at most one particle. 
$\eta_0$ being the same for both density profiles with 
different number of peaks, $N_{\rm peaks}^{(1)}=25$ and $N_{\rm peaks}^{(2)}=16$, allow us to obtain the relation 
$\nu^{(1)}=\left(9+16\nu^{(2)}\right)/25$ between the fraction of vacancies. If we suppose 
that the density profile with 16 peaks has zero vacancies ($\nu^{(2)}=0$) we obtain 
a $36\%$ ($\nu^{(1)}=9/25$) of vacancies for the 25-peaks density profile.

The behavior of particle fluxes during 
the dynamics from 25 to 16 peaks can be seen in Fig. \ref{evolving}. 
Panel (d) shows the $x$-component 
of the flux, $J_x({\bm r},t)$, at the instant $t\approx 0$. The other $y$-component has, by symmetry, exactly 
the same behavior and can be obtained from the $x$-component by a $90^{\circ}$ rotation.  
We can see how the layers close to the soft-walls 
move to the center of the cell (the direction of fluxes of left and right extremal layers 
point to the right and to the left respectively). Moreover the peaks belonging to the intermediate
chain is asymmetrically decomposed by diffusion to  
left and right creating and effective flux to the centre of the cell. The same occurs with the central peak 
which is symmetrically smeared out by diffusion. At further times the density profile 
becomes fluid-like over the whole cell (except for the external layer which keeps certain structure) 
and then it is reconstructed to get a total amount of 16 peaks. 
In panel (f) we show the spatial inhomogeneities of 
$J_x({\bm r},t_2)$ (for $t_2\approx T_{\rm sat}$) close to the equilibrium: During the last steps 
of peaks formation a set of pairs of fluxes of much 
less magnitude coming from both, left and right, directions converge to the 16 particle positions.

\section{C/K nucleation induced by the presence of obstacles}
\label{obstacle}

\begin{figure*}
\epsfig{file=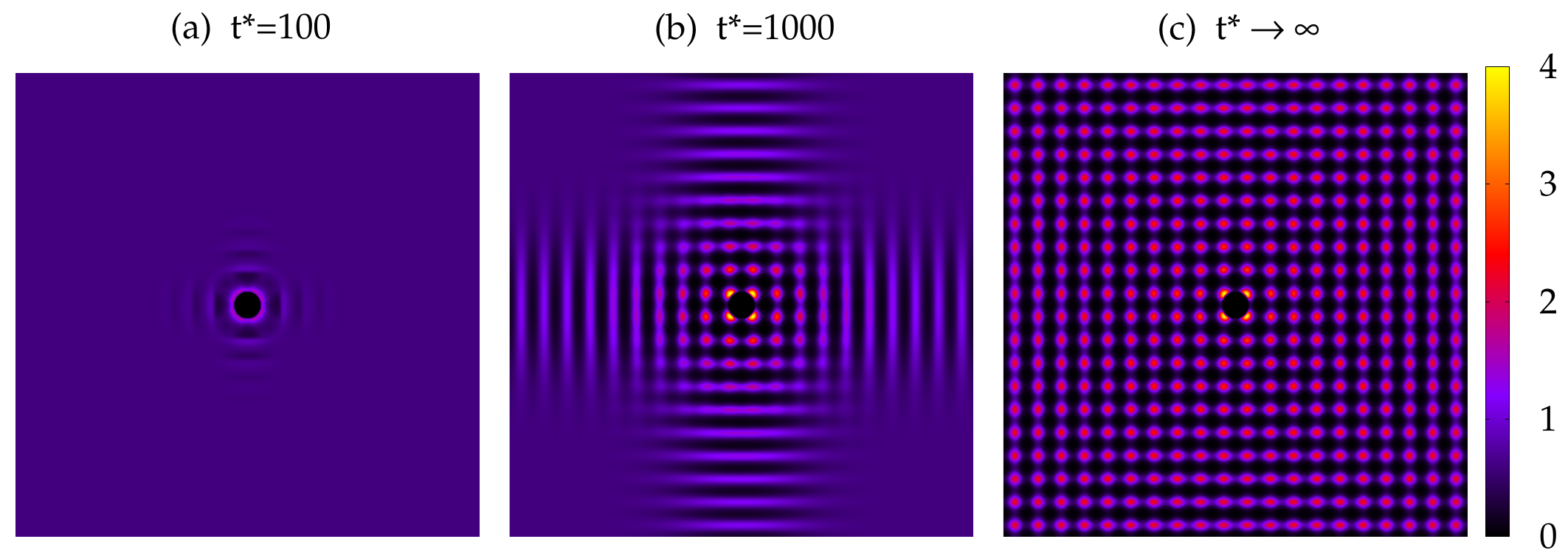,width=6.in}
\epsfig{file=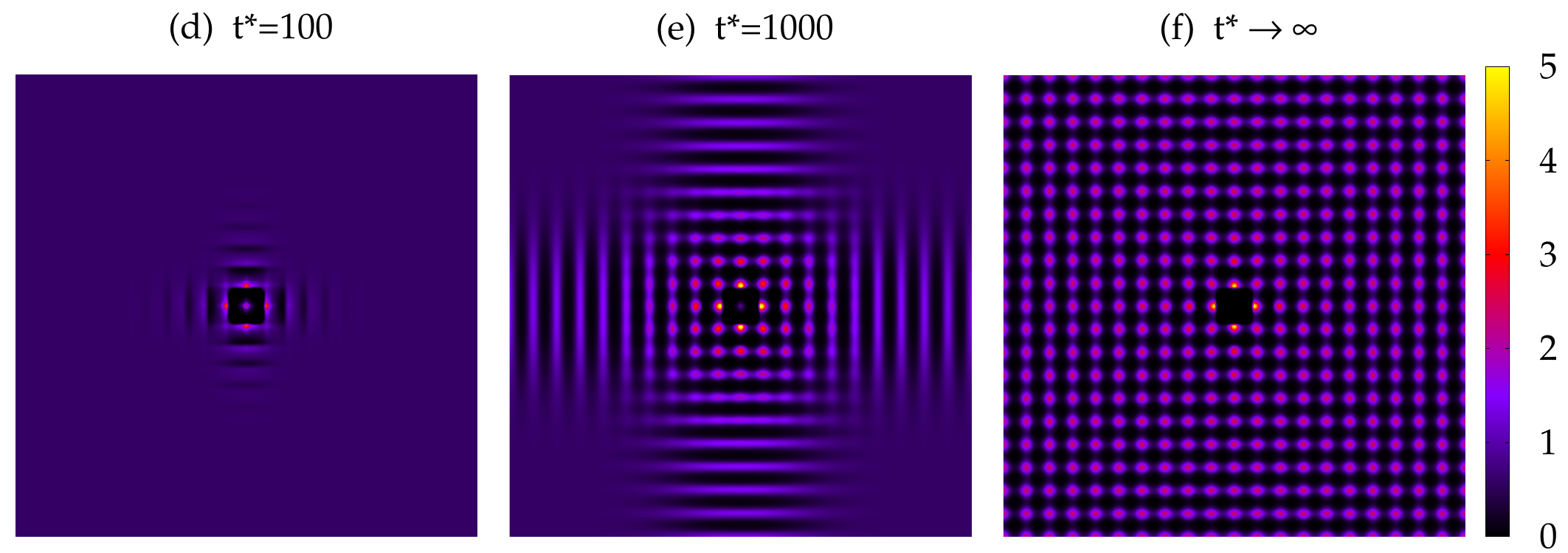,width=6.in}
\caption{(Color online). A sequence of three density profiles $\rho^*({\bm r},t)$
following the dynamics from the DDFT. The profiles correspond to three different times (as labeled)
during the crystallization of PHS around a circular obstacle of diameter $D=\sigma$ (a)-(c) and a square obstacle
of side length and $D=1.5\sigma$ (d)-(f).
The initial density profile was taken constant with packing fraction $\eta_0=0.6$ while periodic
boundary conditions were used.} 
\label{obstacle1}
\end{figure*}

This section is different from the previous ones in one important aspect: the kind of 
external potential used to promote the heterogeneous nucleation of K or C phases. 
We have introduced a strong repulsive potential inside 
a spatial region of circular, square, rectangular or triangular symmetries, with the aim of mimicking 
a hard obstacle at the centre of the box. The obstacle size was chosen to have a few lattice parameters, 
$q\sigma$ ($q\in \mathbb{Q},\ 1\leq q \leq 12$), 
and periodic boundary conditions were used. The size of the square box, $h$, inside which the DDFT equation
is numerically solved was selected large enough to guarantee the correct relaxation of the density profiles 
at long distances from the obstacle. Also, the specific value of $h$ was selected at a local minimum of 
the oscillatory free-energy profile as a function of $h$. 
The main purpose here is to study the dynamics of the heterogeneous nucleation promoted by the presence 
of obstacles with different symmetries. 

First, we use obstacles with different geometries but with the property that they have at least 
fourfold rotational symmetry (i.e. they are invariant under rotations of $90^{\circ}$). 
These are the circular and the square obstacles. In Fig. \ref{obstacle1} we present a sequence of three 
density profiles, $\rho({\bf r},t_i)$ ($t_1<t_2<t_3=T_{\rm sat}$), following the dynamic evolution 
to equilibrium from a constant-density initial condition 
(with $\eta_0=0.6$) and for external potentials of circular (a)-(c) and square symmetries (d)-(f);
these potentials mimic strong repulsive objects of sizes $D=\sigma$ and $1.5\sigma$ (corresponding to the 
values of diameter and side-length, respectively).

The first stages in the dynamics consist of the propagation of four symmetric fronts of C ordering 
along the two perpendicular ($x$ and $y$) directions. These fronts propagate with finite velocity 
from the obstacle to the box boundaries (see Fig. \ref{obstacle1}). 
Obviously the four fronts form an square wave and local maxima of the density profile are located, by 
interference effects, at the corners of the square front. The heterogeneity of the density profile 
along the front induces a secondary mechanism which takes place at longer times: the localization of 
particles by migration along the perimeter of the square front to their final equilibrium locations
at the nodes of a simple square lattice of lattice parameter $a_{\rm K}$ (corresponding 
to a metastable K phase at bulk). Note that, for this density $\eta_0$, the stable phase is C, 
but the obstacle stabilizes the K phase. The dynamics of PHS around an obstacle with circular or 
square symmetries are similar, as can be seen from the figure. The relevant variable that determines 
the final structure of the K phase is the diameter of the obstacle; for a circle with $D=\sigma$, 
panels (a)-(c), values of density peaks in contact with the obstacle are higher than the rest. 
Also, a line joining these peaks outlines the unit cell of the simple square lattice of the metastable 
K phase. For a square obstacle of similar size the structure (not shown) is identical:
the highest density peaks are located at the corners of the square obstacle. By increasing the 
size of the obstacle up to $D=1.5\sigma$ one obtains the final structure shown in panel (f). 
Now the square outlined by the density peaks in contact with the obstacle is larger than the unit cell 
and rotated $45^{\circ}$ with respect to the $x$ axis. 
The presence of the obstacle generates a vacancy of just one particle at its centre while 
for $D=\sigma$ the structure is defect-free. Again the same density profile 
is generated for a circular obstacle of diameter $D=1.5\sigma$.

\begin{figure*}
\epsfig{file=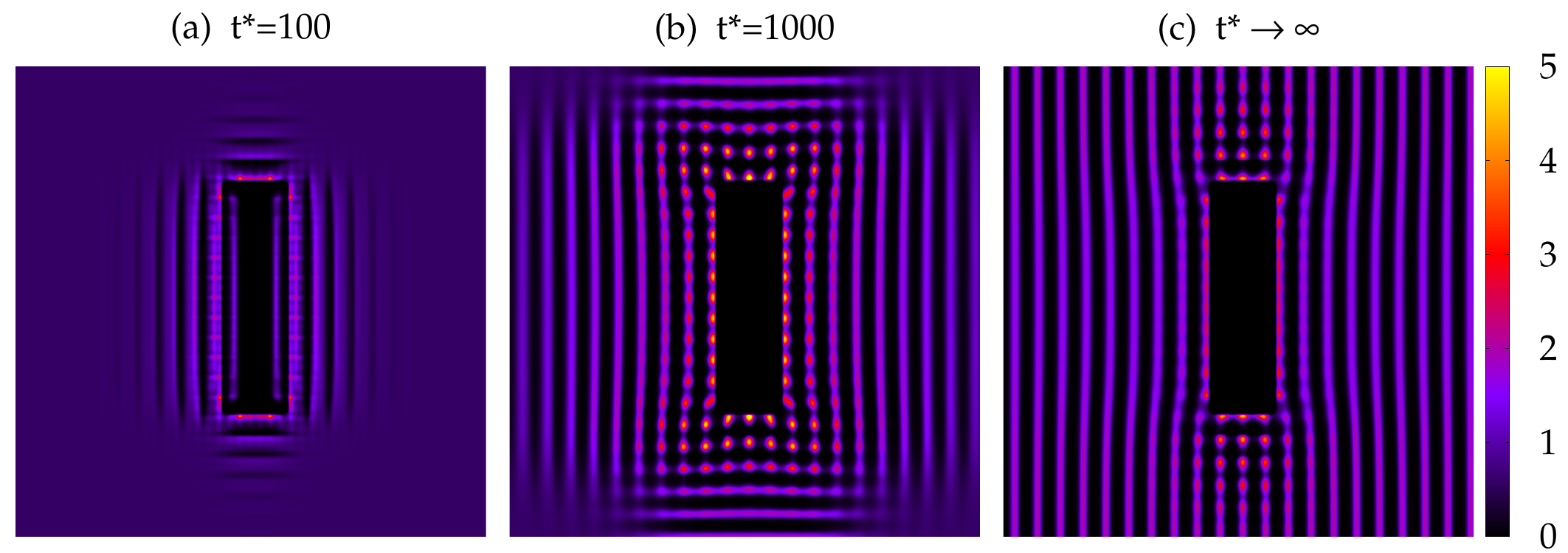,width=6.in}
\epsfig{file=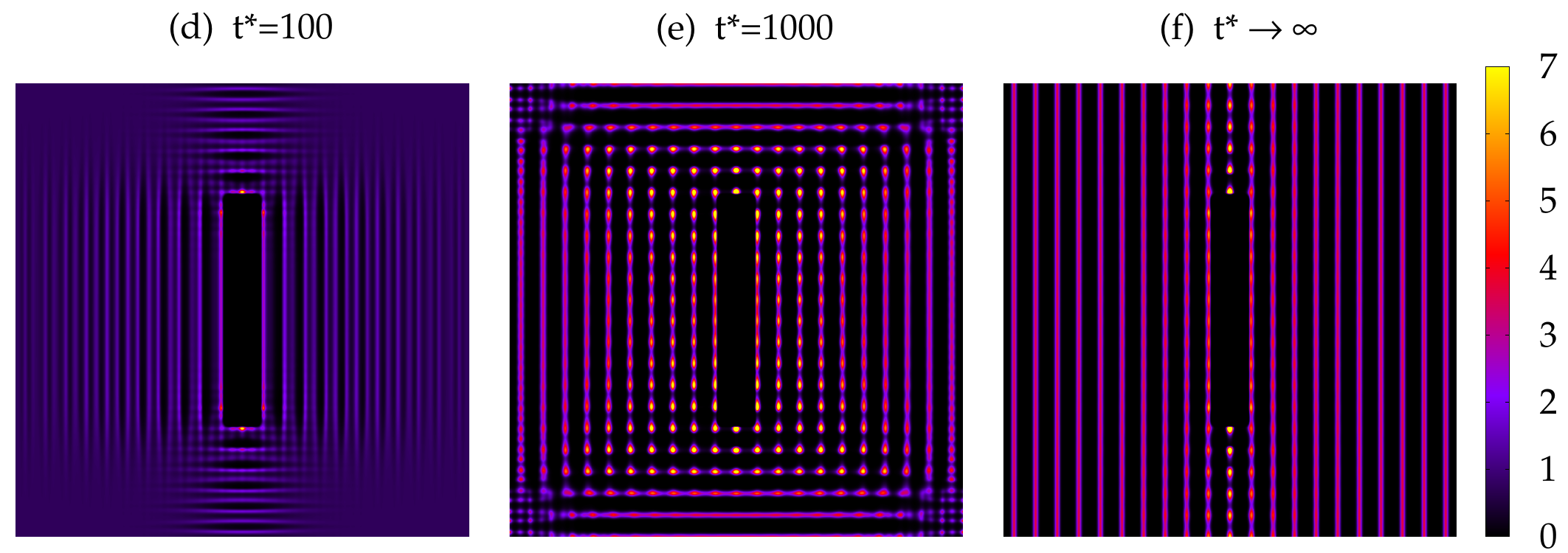,width=6.in}
\caption{(Color online). A sequence of density profiles $\rho^*({\bm r},t)$ corresponding to three different times 
(as labeled) during the columnarization of PHS around a rectangular obstacle of width ($D$) and length ($L$)
equal to $(D,L)=(3.15,12)\sigma$ (a)--(c) and $(D,L)=(1.65,12)\sigma$ (d)--(f). The initial density profiles 
were taken constant with $\eta_0=0.6$ (a)--(c) and $\eta_0=0.75$ (d)--(f).} 
\label{obstacle2}
\end{figure*}

The second study concerns the dynamics of heterogeneous formation of C/K phases around an obstacle 
without the fourfold symmetry. We analyse two obstacles. The first is a rectangle with a long side-length
of $L=12\sigma$, and short side-lengths of $D=3.15\sigma$ and $D= 1.65\sigma$. The other 
is an equilateral triangle of side-length $D=3\sigma$. A sequence of density profiles obtained
during the dynamic evolution, taken at three different times, are shown in Figs. \ref{obstacle2}(a)--(c) 
and (d)--(f) for rectangular obstacles with $D=3.15\sigma$ and $D= 1.65\sigma$, respectively,
and in Fig. \ref{obstacle3} (a)--(c) for the triangular obstacle. As can be seen, the rectangle stabilizes 
a C phase, with columns parallel to the longest side of the rectangle. 
Interestingly, after a time where a rectangular front of C symmetry is propagated from the obstacle, 
a further localization of particles takes place. This localization proceeds by 
particle migration along the perimeter of the front, similar to the cases with obstacles of
circular and square geometries, and extends up to three layers from the obstacle for $D=3.15\sigma$
and to the whole area for $D= 1.65\sigma$. There is however an important difference in this case: after the
second stage, particles again delocalize, restoring the C layers parallel to the 
long side-length. Therefore equilibrium profiles correspond to a defected C phase with disrupted 
columns (three or one layer for $D=3.15\sigma$ and $D=1.65\sigma$, respectively, as shown in 
Fig. \ref{obstacle2}) formed by particles with some degree of localization. 
No perfect commensuration between the difference $h-D$ 
(with $h$ the width of the box) and the lattice parameter $a_{\rm C}$ corresponding to the stable C  phase 
at bulk, as it occurs for $D=3.15\sigma$, generates a deformation of columns around the obstacle [see panel (c)].

\begin{figure}
\epsfig{file=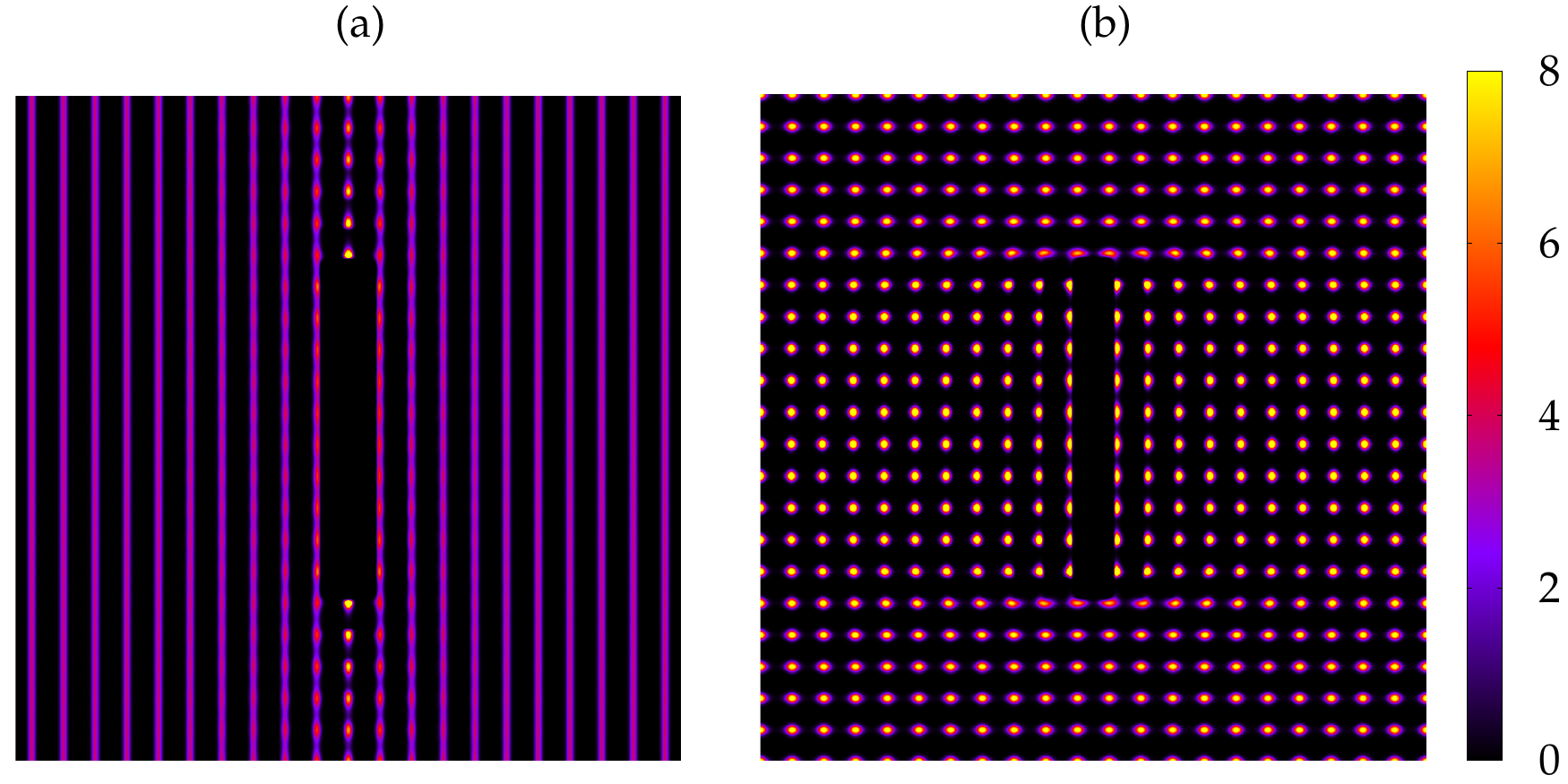,width=3.2in}
\caption{(Color online). Equilibrium density profiles as obtained from the final converged states of the dynamic evolution 
following the DDFT and considering a constant density initial profile with $\eta_0=0.75$. The presence of a rectangular 
obstacle of length $L=12\sigma$ and width $D=1.65\sigma$ (a) and  $1.14\sigma$ (b) are imposed for 
$t\geq 0$ through a corresponding external potential.}
\label{final}
\end{figure}

\begin{figure*}
\epsfig{file=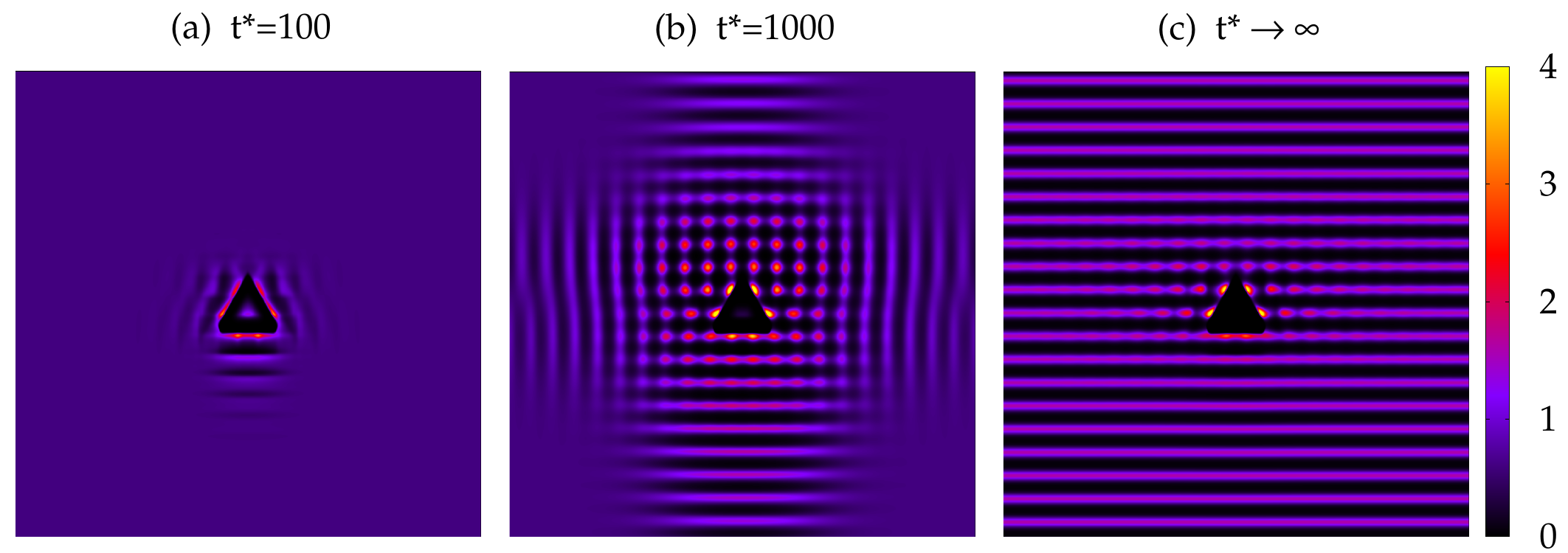,width=6.in}
\epsfig{file=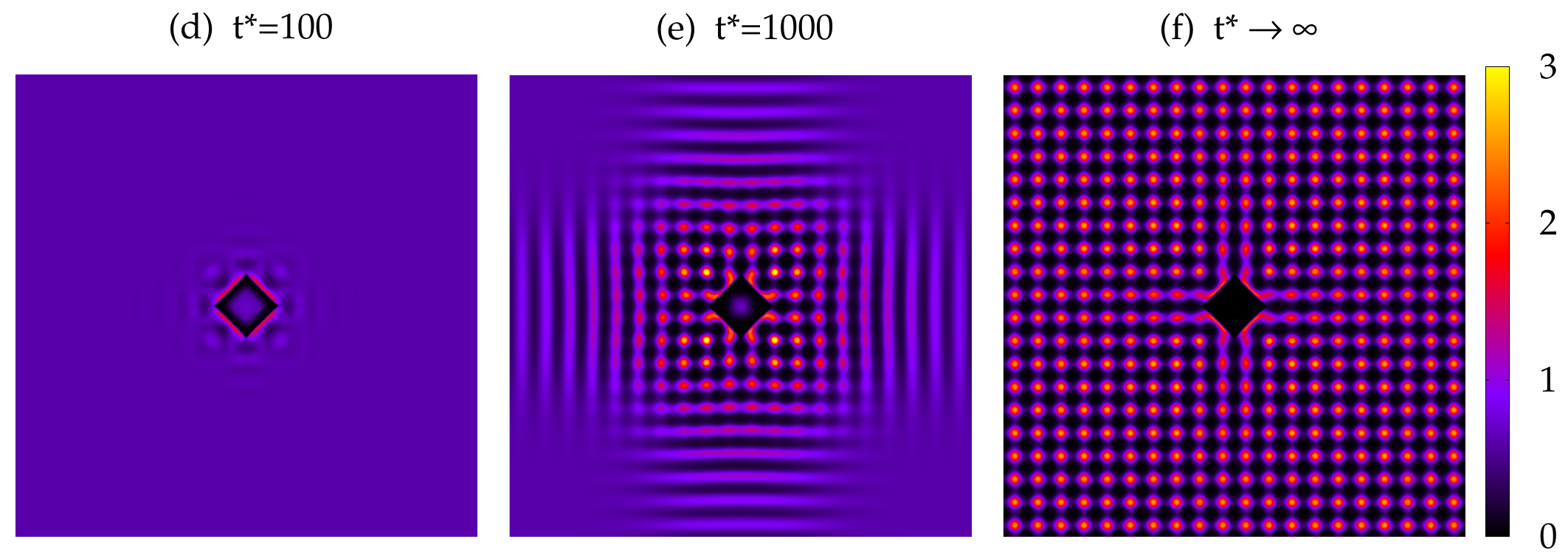,width=6.in}
\caption{(Color online). A sequence of density profiles $\rho^*({\bm r},t)$ corresponding 
to three different times  
(as labeled) during the crystallization of PHS around: (a)-(c) an equilateral triangle of 
side-length $D=3\sigma$ with one of its sides parallel to the $x$-axis, and (d)-(f) a square
rotated by 45$^{\circ}$ and with side length $D=2.6\sigma$. 
The initial density profile was constant, with $\eta_0=0.6$.}
\label{obstacle3}
\end{figure*}

The symmetry of the final equilibrium density profile 
that growths from a rectangular obstacle, considering a constant initial density 
profile with $\eta_0$ slightly above its bulk C-K value, strongly depends on $D$. 
Selecting $D$ well or not well commensurated 
with $a_{\rm C}$ we obtain as $t\to\infty$ density profiles with C or K symmetries respectively as 
they are shown in Fig. \ref{final}.

Another example of an obstacle without the fourfold symmetry is an equilateral triangular obstacle which obviously has 
a $60^{\circ}$ rotational symmetry. We have found that a stable C phase is induced at $\eta_0=0.6$ when one 
of the triangle sides is parallel to $x$ or $y$-Cartesian axes (the same directions of the lattice vectors). 
The set of three density profiles during the evolution dynamic is shown in Fig. \ref{obstacle3} (a)-(c).  
We again see: (i) a first propagation of three triangular C fronts with their further reorientation to 
form four fronts in the two mutual perpendicular directions at larger distances, (ii) a further 
localization of particles around the nodes of a square lattice which extend to few layers from the obstacle, 
and (iii) the final delocalization of particles along the C layers, except for some particles 
in the neighborhood of the contact between the columns and the obstacle. 

To end this section we show the results of the dynamics of heterogeneous crystallization induced by the presence 
of a $45^{\circ}$ rotated square obstacle with respect to the lattice vectors of the K phase. The results are 
shown in Fig. \ref{obstacle3} (d)-(f). Again this is an example of an obstacle with fourfold symmetry and again 
we obtain a dynamic behavior similar to that corresponding to the non-rotated square. The only differences consist 
on the very initial steps of the dynamics evolution when four rhombus-like fronts depart from the obstacle 
while the other difference is related to the particle distribution close to the obstacle at equilibrium. 
See in panel (f) for $t\to\infty$ 
the presence of four fluid-like layers close to the square sides which are deformed to connect the $x$ and $y$ 
lattice directions.

\section{Conclusions}
\label{conclusions}

We have used the DDF formalism, based on the FMT for a fluid of PHS, to study the dynamics of 
heterogeneous nucleation of the K phase when the fluid is confined by soft-repulsive walls. 
The walls define a lattice of periodically spaced square cells that confine the fluid. 
The study is divided into three parts, each corresponding to a specific initial condition: (i) constant 
density profile, (ii) density profile with C symmetry and (iii) density profile with K symmetry. 
We have characterized the dynamics using different quantities, such as 
saturation time, interlayer fluxes and their maximum value and total number of extrema, and total  
(interlayer plus intralayer) fluxes. These quantities are analysed as a function of cell size and
correlated with some features of the equilibrium density profile, such as 
absolute maximum over the cell and total number of layers. 

We found that, for poorly commensurate cells (i.e. with a lattice parameter 
incommensurate with that of a metastable 
K phase at bulk), the structure of the density profile consists of smeared-out peaks with values
lower that those for well-commensurate cavities. In addition, the dynamics is dominated by strong 
interlayer fluxes which expel particles from the walls to the interior of the cavity. 
The equilibrium configuration is reached by further interchange of particles between neighbouring
layers, resulting in moderately localized peaks. 
By contrast, in the case of well-commensurate cells,
intralayer fluxes are dominant, with particles localising at the nodes of the simple square lattice. 
Although interlayer fluxes are lower for well-commensurate cavities, they exhibit a more complex behaviour: 
strong non-monotonicity with presence of a high number of extrema, and damping oscillations which increase
the saturation time before equilibrium is reached. This highly non-linear behaviour strongly correlates 
with longer saturation times, which dramatically increase with number of layers. As a function of cell 
size, this number exhibits a rapid increase for the most noncommensurate cavities (those containing a 
fluid-like density profile). However we have checked that this abrupt increase does not imply a phase 
transition, which is discarded due to the soft character of the external potential.

When the dynamics departs from a C density profile, in most cases the final state is the usual
symmetric K phase. However, for some special cells, in particular those which commensurate with the C 
period at bulk, the equilibrium state is an asymmetric K phase in which layers have a different number of 
peaks along the $x$ and $y$ directions. When this occurs, the free energy of the asymmetric profile is 
lower. Finally, we also used previously converged symmetric K-phase density profiles scaled to the new cell as initial condition. 
We observed the delocalization of 
density peaks and the presence of asymmetric fluxes of particles from the walls to the center, with 
inner layers being the first to melt. Further reconstruction of density peaks from a fluid-like profile 
gives rise to a lower number of layers, but these are well commensurate with cell size.

A final study concerns the dynamics of heterogeneous growth of C or K phases from a obstacle 
with circular, square, triangular or rectangular symmetry. The K phase grows from obstacles with
circular or square symmetries, since they have the same fourfold symmetry. By contrast, when obstacles
do not have fourfold symmetry and they reasonably commensurate with the C-phase lattice parameter,
the final equilibrium state is generally a C phase, with layers parallel 
to the long side length of the rectangle or to one of the triangular sides. However, the dynamics 
in this case is far from simple. Density waves of C symmetry propagate from the obstacle with 
further localization of particles along these fronts; these waves extend to a few layers or even to the 
whole area. Finally, particles localize more strongly, until a 
regular K square lattice is created (for circular and square objects), 
or they delocalize again to recover the C phase, which is the final equilibrium 
state (for rectangular and triangular objects).
   
\acknowledgments

Financial support from MINECO (Spain) under grants FIS2013-47350-C5-1-R and
FIS2015-66523-P are acknowledged.

\end{document}